\documentclass[12pt,preprint]{aastex}

\usepackage{graphicx}
\usepackage{txfonts}
\usepackage{longtable}
\usepackage{lscape}
\usepackage{natbib}

\shorttitle{Polarization signatures of J-state interferences}
\shortauthors{Belluzzi and Trujillo Bueno}

\begin{document}

\title{The impact of quantum interferences between different 
$J$-levels \\
on scattering polarization in spectral lines}

\author{{\sc Luca Belluzzi}\altaffilmark{1,2} {\sc and Javier Trujillo 
Bueno}\altaffilmark{1,2,3}}
\altaffiltext{1}{Instituto de Astrof\'isica de Canarias, E-38205 La Laguna, 
Tenerife, Spain}
\altaffiltext{2}{Departamento de Astrof\'isica, Facultad de F\'isica, 
Universidad de La Laguna, Tenerife, Spain}
\altaffiltext{3}{Consejo Superior de Investigaciones Cient\'ificas, Spain}

\begin{abstract}
The spectral line polarization produced by optically pumped atoms contains a 
wealth of information on the thermal and magnetic structure of a variety of 
astrophysical plasmas, including that of the solar atmosphere.
A correct decoding of such information from the observed Stokes profiles 
requires a clear understanding of the effects that radiatively induced 
quantum interferences (or coherences) between pairs of magnetic sublevels
produce on these observables, in the absence and in the presence of magnetic 
fields of arbitrary strength.
Here we present a detailed theoretical investigation on the role of coherences 
between pairs of sublevels pertaining to different fine-structure $J$-levels, 
clarifying when they can be neglected for facilitating the modeling of the 
linear polarization produced by scattering processes in spectral lines. 
To this end, we apply the quantum theory of spectral line polarization and 
calculate the linear polarization patterns of the radiation scattered at 
90$^{\circ}$ by a slab of stellar atmospheric plasma, taking into account and 
neglecting the above-mentioned quantum interferences. 
Particular attention is given to the $^2S-{^2P}$, $^5S-{^5P}$, and $^3P-{^3S}$ 
multiplets. We point out the observational signatures of this kind of 
interferences and analyze their sensitivity to the energy separation 
between the interfering levels, to the amount of emissivity in the background 
continuum radiation, to lower-level polarization, and to the presence of a 
magnetic field.
Some interesting applications to the following spectral lines are also 
presented: Ca~{\sc ii} H and K, Mg~{\sc ii} h and k, Na~{\sc i} D$_1$ and 
D$_2$, the Ba~{\sc ii} 4554~{\AA} and 4934~{\AA} resonance lines, the 
Cr~{\sc i} triplet at 5207~{\AA}, the O~{\sc i} triplet at 7773~{\AA}, the 
Mg~{\sc i} $b$-lines, and the H$\alpha$ and Ly$\alpha$ lines of H~{\sc i}.
\end{abstract}

\keywords{Polarization - Scattering - Stars: atmospheres - Sun: atmosphere
- Sun: surface magnetism}

\section{Introduction}
Over the last few years, we have witnessed renewed interest in the spectral 
line polarization produced by the presence of population imbalances and 
quantum interferences (or coherences) among the magnetic sublevels of 
atomic energy levels \citep[e.g.,][]{Cas07,JTB09,Ste09,Man11}. 
This so-called atomic level polarization is typically produced by anisotropic 
radiation pumping processes, which are particularly efficient in the outer 
layers of stellar atmospheres, where the depolarizing effect of isotropic
collisions tends to be negligible. 
Particular attention has been given to coherences between pairs 
of magnetic sublevels pertaining to each particular $J$-level of the atomic 
model under consideration ($J$ being the level's total angular momentum).
The sensitivity of these coherences to the Hanle effect produces changes 
in the linear polarization of the emergent spectral line radiation, 
which can be exploited for detecting magnetic fields that are too weak and/or 
too tangled so as to produce measurable Zeeman polarization signals.

In general, quantum interferences are also present between pairs of magnetic 
sublevels pertaining to different $J$-levels. 
It is known that this kind of coherences may produce observable effects in the 
wings of some spectral lines. 
Perhaps the most illustrative solar example is the $Q/I$ pattern observed by 
\citet{Ste80b} across the Ca {\sc ii} H and K lines \citep[see also][]{Gan05}, 
which shows a positive signal at the core of the K line ($J_{\ell}=1/2$ and 
$J_u=3/2$), a sign reversal between the two lines, and zero polarization 
at the center of the H line ($J_{\ell}=1/2$ and $J_u=1/2$). 
That peculiar $Q/I$ pattern could be explained by \citet{Ste80} using a simple 
theoretical model that accounts for the possibility of quantum interferences 
between the two upper $J$-levels of the Ca~{\sc ii} resonance lines.

A rigorous theoretical framework for describing the spectral line polarization 
produced by radiatively induced population imbalances and quantum coherences, 
in the presence of arbitrary magnetic fields, is the density-matrix theory 
described in the monograph ``Polarization in Spectral Lines'' (Landi 
Degl'In\-no\-cen\-ti \& Landolfi 2004, hereafter LL04). 
This theory is based on the hypothesis that the pumping radiation field 
has no spectral structure over frequency intervals larger than the frequency 
separation between the interfering levels (flat-spectrum approximation).
Because of this approximation, the theory is very suitable for treating 
spectral lines that can be described under the hypothesis of complete frequency 
redistribution (CRD), while it cannot account for the effects of partial 
redistribution in frequency (PRD).
Despite of this limitation, this theory represents the most robust quantum 
approach to the physics of polarization developed so far. 
In particular, it accounts for the role of quantum interferences in a very 
general, self-consistent way \citep[e.g., the review by][]{Bel11}.
As a matter of fact, although the Ca~{\sc ii} H and K lines are 35~{\AA} apart 
(the flat-spectrum approximation thus appearing rather restrictive), the theory 
explains very well the sign reversal observed in the blue wing of the H line 
in terms of coherences between the two upper levels of these lines (see LL04). 
Theoretical approaches aimed at including PRD effects in the presence of 
$J$-state interferences can be found in \citet{Lan97} and \citet{Smi11}.

It is known that interferences between pairs of magnetic sublevels pertaining 
to different $J$-levels are the smaller the larger the energy separation 
between them. When this separation is large, their effects are negligible 
in the core of the lines, becoming more important in the far wings 
\citep[see][LL04]{Ste80}.
On the other hand, far from line center the line emissivity is very low, 
and the presence of the continuum generally masks the effects of such
interferences.
Because of these basic arguments, interferences between different $J$-levels
are usually neglected when investigating the scattering polarization
properties of many spectral lines, with a considerable simplification
of the problem. However, this kind of qualitative considerations should not
be applied as rigorous rules, as testified by the Ca~{\sc ii} H and K 
lines and by several other signals of the linearly-polarized solar limb 
spectrum (or second solar spectrum) which show the signatures of this 
kind of interferences \citep[see][]{Gan00,Gan02,Gan05}.
As far as the solar Ca~{\sc ii} H and K lines are considered, it is now clear 
that the reason why quantum interferences between such significantly 
separated lines produce observable effects is because these strong
lines have very extended wings, so that the continuum is not immediately 
reached when moving away from line center, and because the large optical depth 
of the solar atmosphere compensates for the low line emissivity at the 
wavelengths where that observational signature appears 
\citep[cf.][LL04]{Ste80}.

In this paper, we present a systematic theoretical investigation of the role of 
quantum interferences between different $J$-levels.  We aim at clarifying
their observable effects in various interesting multiplets and at providing a 
series of general criteria for establishing under which circumstances their 
effects are expected to be observable, and when, on the contrary, the modeling 
of spectropolarimetric observations can be safely carried out ignoring their 
contribution.
The investigation is carried out within the framework of the above-mentioned 
quantum theory of polarization. 
After briefly introducing the density-matrix formalism (Section~2), and 
presenting the scattering polarization model that is applied for our 
investigation (Section~3), we start focusing the attention on the $^2S-{^2P}$ 
multiplet, the simplest one where interferences between different $J$-levels 
occur (Section~4).
We show that although the large energy separation generally present 
between different $J$-levels makes this kind of interferences usually 
very small, as far as fractional polarization signals (i.e., ratios, such as 
$Q/I$) are considered, their effects are not necessarily negligible.
Indeed, when only line processes are considered (no continuum), the
fractional polarization patterns obtained taking into account and neglecting 
these interferences coincide within a small spectral interval around the
center of the lines, but are found to be very different at all other 
wavelengths, even when the $J$-levels are very separated from each other 
(see Section~4.2).
Particular attention is given to the analysis of the effects of the Doppler 
broadening of the lines, a physical aspect that is found to play an
important role for establishing where and when such interferences can be
neglected.

As previously mentioned, what makes the effects of interferences between
different $J$-levels vanish in the observed polarization patterns is the
continuum, which starts dominating over the line emissivity moving
from the line center to the wings. In Section~4.3, we show that the amount
of continuum needed to mask the signatures of such interferences is the larger 
the smaller the separation between the $J$-levels. 
The effects of interferences are thus expected to be observable either when 
the interfering lines have very extended wings, so that the continuum is not
immediately reached between them (as in the case of the Ca~{\sc ii} H
and K lines), or when the separation between the $J$-levels is sufficiently
small but, as we will see, still large when compared to the Doppler width of 
the corresponding spectral lines.

In the first part of the paper, we analyze the scattering polarization 
profiles of hypothetical multiplets characterized by different values of the 
energy separation between the interfering $J$-levels, of the Doppler width, 
and of the continuum intensity.
The results of this analysis, and their consequences for practical 
applications, are then investigated in detail on various multiplets of 
particular interest for the diagnostics of the magnetism of the solar 
atmosphere (Sections~4.5, 5, 6, and 7).
The effects due to the presence of magnetic fields of various intensities and
configurations are also investigated (Section~8).

\section{Statistical equilibrium equations in the density-matrix formalism}
\label{Sect:theory}
A convenient way to describe the populations of the various magnetic sublevels 
of an atomic system, as well as the quantum interferences between pairs of 
them, is through the matrix elements of the density operator.
The density operator is a very useful theoretical tool for describing any 
physical system which is in a statistical mixture of states 
\citep[e.g.,][]{Fan57}.
If $p_1$, $p_2$, \ldots, $p_i$, \ldots, are the probabilities for a given 
physical system of being in the dynamical states represented by the vectors 
$|\,1>$, $|\,2>$, \ldots, $|\,i>$, \ldots, respectively, the corresponding 
density operator is defined by 
\begin{equation}
	\rho = \sum_i p_i \, |\,i><i\,| \;\; .
\end{equation}
As any other quantum operator, also the density operator is completely 
specified once its matrix elements, evaluated on a given basis of the Hilbert 
space associated with the physical system, are known. 
The matrix elements of the density operator contain all the accessible 
information about the system. 

On the basis of the atomic energy eigenvectors $\{|\, m \! >\}$, the 
matrix elements of the density operator are given by
\begin{equation}
	< m \, |\, \rho \,| \, m^{\prime} \!> \, \equiv \, 
	\rho_{m m^{\prime}} \;\; ;
\end{equation}
the diagonal elements represent the populations of the various energy 
levels, while the off-diagonal elements represent the quantum interferences 
between pairs of them. 
On this basis, the statistical equilibrium equations (SEEs) for the atomic 
density matrix are given by (see Equation~(6.62) of LL04)
\begin{eqnarray}
\label{Eq:SEE}
\frac{\mathrm{d}}{\mathrm{d}t} \, \rho_{mm^{\prime}} & \!\!\!\! = \!\!\!\! & 
-2 \pi \mathrm{i}
\nu_{mm^{\prime}} \rho_{mm^{\prime}} + \sum_{nn^{\prime}} \rho_{nn^{\prime}} 
\,T_{A}(m,m^{\prime},n,n^{\prime}) \nonumber \\
 & & + \sum_{pp^{\prime}} \rho_{pp^{\prime}} \,T_{E}(m,m^{\prime},p,p^{\prime})
+ \sum_{pp^{\prime}} \rho_{pp^{\prime}} \,T_{S}(m,m^{\prime},p,p^{\prime})
\nonumber \\
 & & - \sum_{m^{\prime \prime}} \Big[ \rho_{mm^{\prime \prime}} \,R_{A}(m,
m^{\prime},m^{\prime \prime}) + \rho_{m^{\prime \prime}m^{\prime}} \,R_{A}
(m^{\prime},m^{\prime \prime},m) \Big] \nonumber \\
 & & - \sum_{m^{\prime \prime}} \Big[ \rho_{mm^{\prime \prime}} \,R_{E}
(m^{\prime \prime},m,m^{\prime}) + \rho_{m^{\prime \prime}m^{\prime}}
\,R_{E} (m,m^{\prime},m^{\prime \prime}) \Big] \nonumber \\
 & & - \sum_{m^{\prime \prime}} \Big[ \rho_{mm^{\prime \prime}} \,R_{S}
(m^{\prime \prime},m,m^{\prime}) + \rho_{m^{\prime \prime}m^{\prime}}
\,R_{S} (m,m^{\prime},m^{\prime \prime}) \Big] \;\; .
\end{eqnarray}
These equations describe the transfer ($T$ rates) and relaxation ($R$ rates) of 
populations and coherences due to absorption (index $A$), spontaneous emission 
(index $E$), and stimulated emission (index $S$) processes.
They also describe the effects due to the presence of a magnetic field.
A complete derivation of these equations, as well as the explicit expression 
of the various rates can be found in LL04.

Here we focus our attention on the first term in the right-hand side of 
Equations~(\ref{Eq:SEE}).
This term, proportional to the Bohr frequency $\nu_{mm^{\prime}} = (E_m - 
E_{m^{\prime}})/h$, with $E_i$ the energy of level $|\,i>$ and $h$ the Planck 
constant, is zero both for populations (i.e., for the diagonal elements 
$\rho_{mm}$) and for coherences between degenerate levels, while it 
produces a relaxation of coherences between pairs of non-degenerate levels.
The coherence $\rho_{mm^{\prime}}$ is thus the smaller, the larger the energy 
separation between the levels $|\, m \! >$ and $|\, m^{\prime} \! >$. 
Modifying the frequency separations $\nu_{mm^{\prime}}$ among the various 
magnetic sublevels, a magnetic field modulates, through this term, the 
corresponding coherences $\rho_{mm^{\prime}}$, and consequently the 
polarization of the emitted radiation. 
This is the basic physical mechanism at the origin of the so-called Hanle 
effect.

Finally, it is important to recall that Equations~(\ref{Eq:SEE}) are valid 
under the so-called flat-spectrum approximation. 
This approximation requires the pumping radiation field to be spectrally flat 
over frequency intervals larger than the Bohr frequency relating pairs of 
levels between which quantum interferences are considered, and 
larger than the inverse lifetime of the same levels.

For an atomic system devoid of hyperfine structure (HFS), the matrix elements 
of the density operator are often defined on the basis of the eigenvectors of 
the angular momentum\footnote{We recall that in the presence of intense 
magnetic fields, in the so-called Paschen-Back effect regime, this is not the 
basis of the energy eigenvectors.} \mbox{$\{|\, \alpha J M>\}$}, with $J$ and 
$M$ the quantum numbers associated with the total angular momentum and to its 
projection along the quantization axis, and $\alpha$ a set of inner quantum 
numbers associated with the atomic Hamiltonian\footnote{For an atomic system 
described by the $L-S$ coupling scheme, $\alpha$ could represent, for 
example, the set of quantum numbers ($\beta$, $L$, $S$) which describe the 
electronic configuration, the total orbital angular momentum, and the total 
electronic spin.}.
On this basis, the matrix elements of the density operator are 
\begin{equation}
	< \alpha J M |\, \rho \,| \, \alpha^{\prime} J^{\prime} M^{\prime} > 
	\, \equiv \, \rho \, (\alpha J M,\alpha^{\prime} J^{\prime} M^{\prime}) 
	\;\; .
\end{equation}
The diagonal elements represent the populations of the various magnetic 
sublevels, while the off-diagonal elements the quantum interferences between 
pairs of them.

An atomic model accounting for coherences between pairs of magnetic sublevels 
pertaining either to the same $J$-level or to different $J$-levels within the 
same term is generally referred to as ``multi-term atom'' (see Section~7.5 of 
LL04).
In this case, the flat-spectrum approximation requires the radiation field 
incident on the atom to be flat over frequency intervals larger than the 
frequency separation among the $J$-levels belonging to each term. 
This implies that the incident radiation field must be flat across the 
frequency interval covered by the transitions of a given multiplet.

When coherences between different $J$-levels are neglected, the corresponding 
atomic model is generally referred to as ``multi-level atom'' (see Section~7.1 
of LL04). 
In this case, the flat-spectrum approximation is less restrictive, since it 
requires the radiation field to be constant across frequency intervals 
larger than the Zeeman splitting among the various magnetic sublevels 
pertaining to each $J$-level, and larger than their inverse lifetimes.
In a multi-level atom, the incident radiation field can thus vary across the 
frequency interval spanned by the transitions of a given multiplet.

The SEEs equations and the expressions of the radiative transfer coefficients
for a multi-term and a multi-level atom can be found in Chapter 7 of LL04.

\section{The scattering polarization model}
\label{Sect:model}
In order to analyze the effects of interferences between different $J$-levels,
we consider an atomic system composed of two terms (each being characterized 
by the total orbital angular momentum $L$, by the total electronic spin $S$, 
and by given fine-structure (FS) splittings of the various $J$-levels), and we 
compare the polarization patterns of the radiation emitted across the 
transitions of the corresponding multiplet as obtained both taking into account 
and neglecting such coherences (i.e., as calculated within the framework of a 
two-term atomic model, which accounts for such interferences, and within the 
framework of the corresponding multi-level atomic model, which neglects them).
Although in the multi-level atom case the pumping radiation field can be 
different at the frequencies of the various transitions of the multiplet, in 
order to analyze the net effect of the coherences between different $J$-levels, 
we consider the same incident field, flat across the whole frequency interval 
of the multiplet, both in the case of the two-term atom, and in the case of the 
corresponding multi-level atom.

Interferences between different $J$-levels have a double role: on one hand they 
enter the SEEs and therefore, in principle, their presence may modify 
populations and interferences within the same $J$-level with respect to the 
case in which they are neglected; on the other hand they contribute (exactly 
as populations and interferences within the same $J$-level do) to the radiative 
transfer coefficients.

We focus our attention on the radiation scattered at $90^{\circ}$ by a 
plane-parallel slab of plasma illuminated by the solar continuum radiation 
field. 
In this simple scenario, the fractional polarization of the scattered radiation 
can be calculated through the approximate formula \citep[see][]{JTB03}
\begin{equation}
\label{Eq:emer-pol}
	\frac{X}{I} \approx \frac{\varepsilon^{\, \ell}_X}
	{\varepsilon^{\, \ell}_I+\varepsilon^{\, c}_I} -
	\frac{\eta^{\, \ell}_X}{\eta^{\, \ell}_I+\eta^{\, c}_I} \;\; ,
	\;\;\; {\rm with} \;\; X=Q,U,V \; ,
\end{equation}
where $\varepsilon^{\, \ell}_i$ and $\eta^{\, \ell}_i$ ($i=I,Q,U,V$) are the 
line emission and absorption coefficients, respectively, in the four Stokes 
parameters, while $\varepsilon^{\, c}_I$ and $\eta^{\, c}_I$ are the continuum 
intensity emission and absorption coefficients, respectively.
The incident continuum radiation is assumed to have axial symmetry around the 
normal to the slab (the local vertical), to be unpolarized, and to be flat over 
the frequency intervals covered by the multiplets that will be considered in 
this investigation.
The reference direction for positive $Q$ is assumed perpendicular to the 
scattering plane.
The Doppler width used in the calculation of the line emission and absorption 
coefficients is derived from the temperature and microturbulent velocity of a 
solar model atmosphere at the height where the line-center optical depth is 
unity, for an observation at $\mu=0.1$. 
The semi-empirical atmospheric model $FALC$ of \citet{Fon93} has been used.

The first step is to write down and solve the SEEs for the given incident 
(pumping) continuum radiation field.
We describe the incident radiation through the radiation field tensor 
$J^K_Q(\nu)$ (see Equation~(5.157) of LL04 for its definition), taking the 
quantization axis along the local vertical.
Due to the cylindrical symmetry of the incident continuum radiation around this 
direction, only two components of the radiation field tensor are non-zero: 
$J^0_0$ and $J^2_0$.
The former describes the average intensity of the radiation field over all the 
directions of propagation, while the latter gives a measure of the anisotropy 
of the radiation field. 
Under such circumstances, it is customary to describe the radiation field 
through two equivalent dimensionless quantities, the mean number of photons
per mode ($\bar{n}$) and the anisotropy factor ($w$), defined by (LL04)
\begin{equation}
	\bar{n}=\frac{c^2}{2 h \nu^3} J^0_0 \;\; , 
	\;\;\;\;\; 
	w=\sqrt{2} \frac{J^2_0}{J^0_0} \;\; .
	\label{Eq:w-barn}
\end{equation}

In the absence of magnetic fields and of other mechanisms able to break 
the cylindrical symmetry of the problem, if the quantization axis is taken 
along the symmetry axis (as we are assuming here), the only interferences that 
can be excited are those between pairs of magnetic sublevels pertaining to 
different $J$-levels and characterized by the same value of the magnetic 
quantum number $M$.
These are exactly the interferences that we are going to investigate in the 
following sections.

Once the SEEs are solved, and the density-matrix elements are known, we 
calculate the line emission and absorption coefficients appearing in 
Equation~(\ref{Eq:emer-pol}), according to the expressions given in LL04 
(Chapter 7).
The continuum emission coefficient $\varepsilon^{\, c}_I$ will be a free 
parameter of the problem, used to analyze the effect of the continuum in 
masking the observational signatures of the coherences under investigation.
The value of the continuum absorption coefficient is calculated here
through the equation
\begin{equation}
	\eta_I^{\, c} = \frac{\varepsilon^{\, c}_I}{B(T)} \;\; ,
	\label{Eq:etaIc}
\end{equation}
where $B(T)$ is the Planck function at the wavelength of the transition 
under investigation and at the temperature chosen for the calculation of the 
Doppler width of each line, as previously discussed.

\section{The $^2S-{^2P}$ multiplet}
We start our investigation considering the $^2S-{^2P}$ multiplet, the 
simplest one where interferences between different $J$-levels occur, and one of 
the most interesting given the large number of strong spectral lines belonging 
to this multiplet that are observed on the Sun (Ca~{\sc ii} H and K, 
Mg~{\sc ii} h and k, Na~{\sc i} D$_1$ and D$_2$, Ly$\alpha$, etc.).

This multiplet consists of the following two transitions: 
$J_{\ell}\!=\!1/2 \rightarrow J_u\!=\!1/2$ (transition of H or D$_1$ type, in 
the following referred to as transition 1) and 
$J_{\ell}\!=\!1/2 \rightarrow J_u\!=\!3/2$ (transition of K or D$_2$ type, in 
the following referred to as transition 2).
We want to investigate the effects of the quantum interferences between the 
upper levels $J_u\!=\!1/2$ and $J_u\!=\!3/2$.
Since the common lower level of the transitions of this multiplet, having 
$J_{\ell}\!=\!1/2$, cannot be polarized by the unpolarized incident radiation
field, the SEEs considerably simplify with respect to the general case. 
Moreover, $\eta_Q^{\, \ell}=0$ so that the second term in the right-hand side 
of Equation~(\ref{Eq:emer-pol}), which describes the effects of 
dichroism, does not bring any contribution to the emergent polarization.
If stimulation effects are neglected (which is a good approximation in the 
solar atmosphere), and no magnetic fields are considered, it is possible to 
find an analytical solution of the SEEs, and rather simple analytical 
expressions for the ratio $p_Q=\varepsilon^{\, \ell}_Q/\varepsilon^{\, \ell}_I$,
both in the case in which the interferences between the two upper $J$-levels 
are taken into account, and in the case in which they are neglected (see 
Equations~(\ref{Eq:eps3Lp}) and (\ref{Eq:eps2Tp}) in the Appendix).

\begin{figure}[!t]
\includegraphics[width=\textwidth]{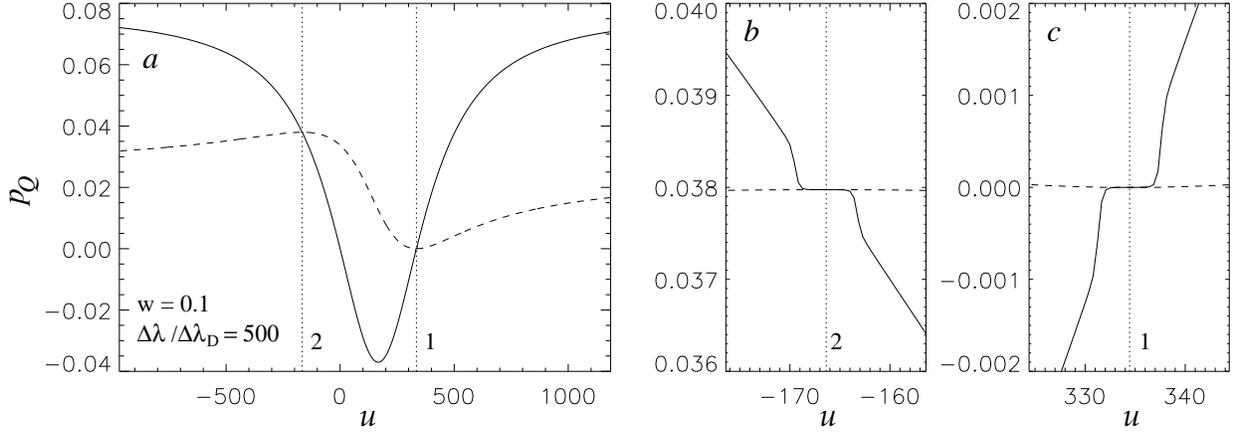}
\caption{\footnotesize{Panel $a$: $p_Q$ profiles obtained taking into account 
(solid line) and neglecting (dashed line) interferences between the upper 
$J$-levels, plotted as a function of the reduced wavelength 
$u=(\lambda-\lambda_0)/\Delta \lambda_D$. 
The results refer to a hypothetical $^2S-{^2P}$ multiplet with a wavelength 
separation between the two FS components equal to 500 times the Doppler width
of the lines (the same Doppler width has been assumed for the two lines).
The vertical dotted lines indicate the position of the two FS components.
The reference wavelength ($\lambda_0$) is the one corresponding to the energy
difference between the centers of gravity of the two terms.
All the calculations have been performed assuming the $L-S$ coupling scheme 
to hold.
Panel $b$: detail of the ``plateau'' (see the text) shown by the $p_Q$ profiles 
around transition 2.
Panel $c$: detail of the ``plateau'' shown by the $p_Q$ profiles around 
transition 1.}}
\label{Fig:pq-gen}
\end{figure}

The $p_Q$ profiles obtained taking into account (solid line) and neglecting 
(dashed line) interferences between the upper $J$-levels, assuming $w\!=\!0.1$ 
and a wavelength separation between the two components
$\Delta \lambda = 500 \, \Delta \lambda_D$, with $\Delta \lambda_D$ the Doppler 
width of the two lines, are plotted in panel $a$ of Figure~\ref{Fig:pq-gen}.
As seen in the figure, the two profiles coincide in the core of the two lines, 
while they are very different at all the other wavelengths 
\citep[cf.][LL04]{Ste80}.
The most remarkable difference is the sign reversal between the two lines 
shown by the $p_Q$ profile calculated taking into account interferences.
This particular signature of the quantum interferences between the two upper 
$J$-levels has been clearly observed between the H and K lines of Ca~{\sc ii} 
\citep[see][]{Ste80}.
The negligible contribution of interferences between different $J$-levels in 
the center of the single lines of an arbitrary multiplet has already been 
discussed in Section~10.17 of LL04, where an analytical expression for the 
value of $p_Q$ in the core of the lines is derived (cf. 
Equation~(\ref{Eq:pq-2Tcore})).

As discussed in LL04, under the hypotheses previously introduced, the 
asymptotic value of $p_Q$ for an arbitrary two-term atom is equal to the 
(constant) $p_Q$ value of a two-level atom with $J_u = L_u$ and 
$J_{\ell} = L_{\ell}$. The analytical expression of $p_Q$ for a two-level atom 
can be found in the Appendix (Equation~(\ref{Eq:pq-2L})).

\subsection{Dependence on $\bar{n}$ and $w$}
\label{Sect:barn-w}
\begin{figure}[!t]
\centering
\includegraphics[width=0.5\textwidth]{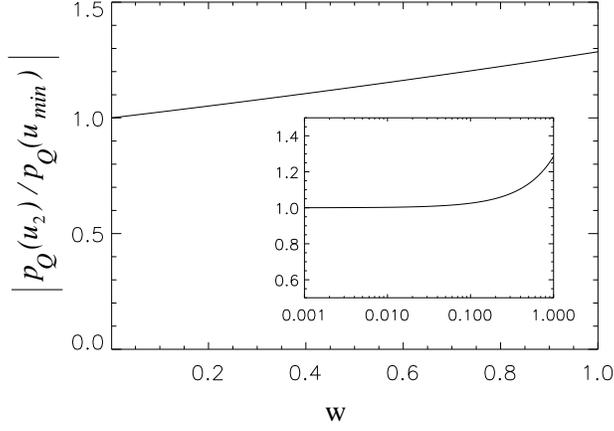}
\caption{\footnotesize{Absolute value of the ratio between the values of 
$p_Q$, calculated taking into account interferences, at the wavelength position 
of transition 2, and at the wavelength position of the negative minimum between 
transitions 2 and 1, plotted as a function of the anisotropy factor (the 
interval between 0 and 1 has been considered).
The inset graphic shows the same plot with the abscissa in logarithmic scale.}}
\label{Fig:w-dep}
\end{figure}

As shown by Equations~(\ref{Eq:eps3Lp}) and (\ref{Eq:eps2Tp}), the quantity 
$p_Q$, as calculated both taking into account and neglecting interferences 
between different $J$-levels, depends on the value of $w$ but not on the value 
of $\bar{n}$.\footnote{Note 
that this result is correct provided that $\bar{n} << 1$, as implicit in 
Equations~(\ref{Eq:eps3Lp}) and (\ref{Eq:eps2Tp}) which are derived neglecting 
stimulation effects.}
Concerning the dependence on the anisotropy factor, it should be observed that 
for small values of $w$ the second term in the denominator of 
Equations~(\ref{Eq:eps3Lp}) and (\ref{Eq:eps2Tp}) can be neglected with respect 
to the first one, so that $w$ represents just a scaling factor of the whole 
profiles.
This property can be clearly appreciated from Figure~\ref{Fig:w-dep}, where the 
absolute value of the ratio between the values of $p_Q$ (calculated taking into 
account interferences) at line center of transition 2 and at the 
wavelength position of the (negative) minimum, is plotted as a function of the 
anisotropy factor.
The value of $p_Q$ at line center of transition 2 is given by 
Equation~(\ref{Eq:pq-2Tcore}), while it is not possible to find a simple 
analytical expression for the amplitude of the negative dip.
As it can be observed from the figure, the above-mentioned ratio is constant 
and very close to unity for values of $w$ smaller than 0.1. 
For these values, which are those that are commonly encountered in the solar 
atmosphere, the anisotropy factor is just a scaling factor of the whole $p_Q$ 
profile.
Numerical calculations show that the position of the minimum does not depend 
on the particular value of $w$.
If not explicitly stated, all the results presented throughout this 
investigation are obtained assuming $w=0.1$ and $\bar{n}=10^{-3}$.

\subsection{Dependence on the wavelength separation between the two components
and the effect of a finite Doppler width}
\label{Sect:DW-effect}
As shown in panels $b$ and $c$ of Figure~\ref{Fig:pq-gen}, at the wavelength 
positions of the two transitions, the $p_Q$ profiles obtained taking into 
account and neglecting interferences coincide and are constant over a 
wavelength interval of about five Doppler widths.
The mathematical and physical origin of these ``plateaux'' is discussed in 
the Appendix, where it is shown that their boundaries are determined by the 
wavelength positions at which the behavior of the corresponding Voigt profiles 
changes from Gaussian to Lorentzian.

\begin{figure}[!t]
\centering
\includegraphics[width=0.99\textwidth]{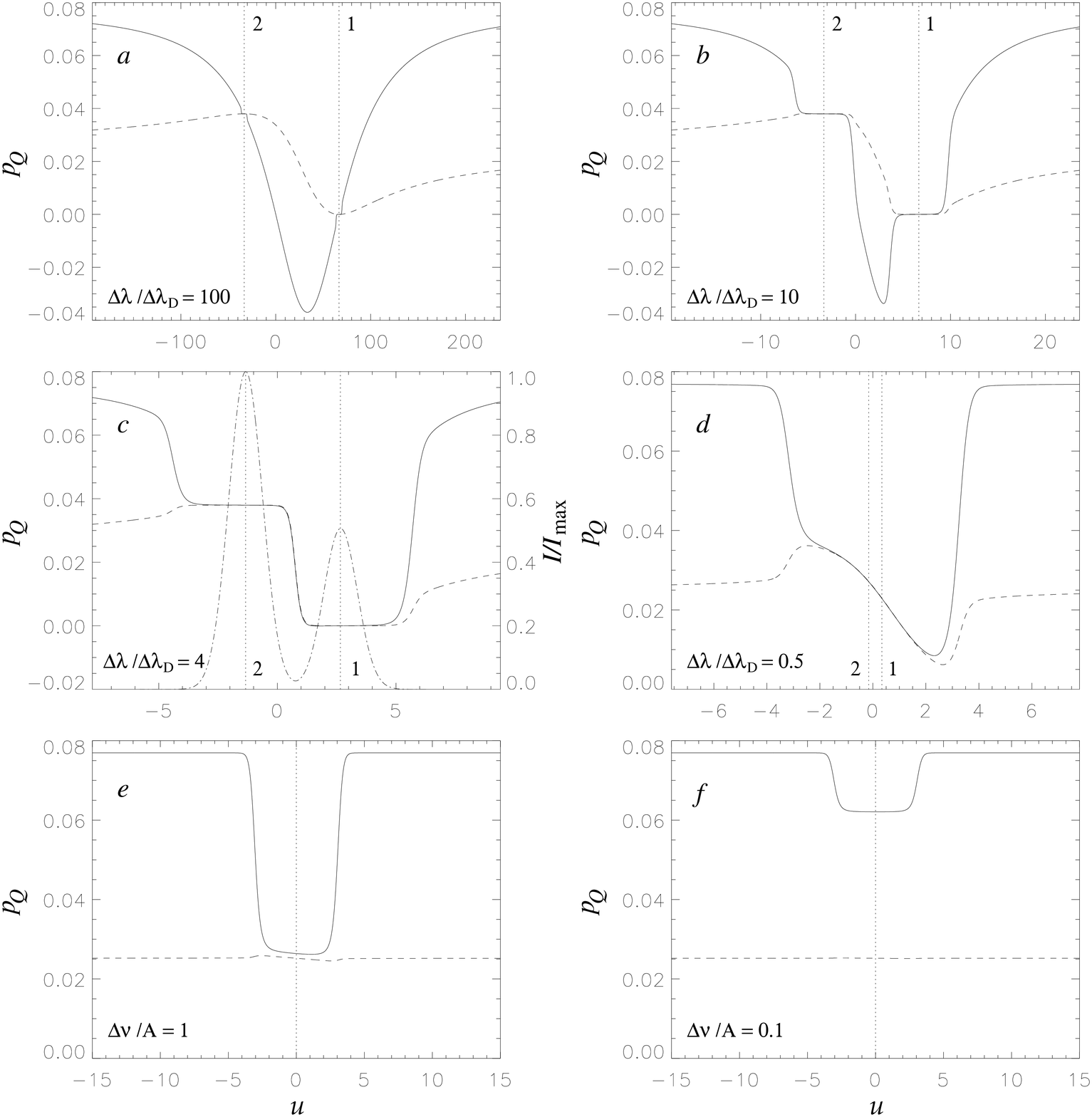}
\caption{\footnotesize{Each panel shows the $p_Q$ profiles obtained taking into 
account (solid line) and neglecting (dashed line) interferences between the 
upper $J$-levels for decreasing values of the separation $\Delta \lambda$ 
between the two components of the multiplet. 
The profiles are plotted as a function of the reduced wavelength $u$.
The reference wavelength is the one corresponding to the energy difference 
between the centers of gravity of the two terms.
The vertical dotted lines indicate the wavelength position of the two 
components (in panels $e$ and $f$ they cannot be distinguished).
In panel $c$ also the intensity profiles of the two lines (normalized to the 
maximum value of $\varepsilon_I^{\, \ell}$ over the whole multiplet) are 
plotted as a function of the reduced wavelength (dash-dotted line), according 
to the scale shown on the right ordinate axis.}}
\label{Fig:Dl-dep}
\end{figure}

Since the extension of the plateaux does not show significant variations with 
the wavelength separation between the two components of the multiplet, their
presence only marginally affects the $p_Q$ profile when the separation between 
the two components is much larger than the Doppler width of the two lines (see 
panel $a$ of Figures~\ref{Fig:pq-gen} and \ref{Fig:Dl-dep}), while it starts 
modifying the overall shape of the $p_Q$ profiles when the distance between the 
two components is of the same order of magnitude as the Doppler width of the 
lines (see panel $b$ of Figure~\ref{Fig:Dl-dep}).

When the wavelength separation between the two components is equal or smaller 
than the extension of the two plateaux, these start merging, so that between 
the two lines all the signatures of the interferences between the upper 
$J$-levels (such as the negativity previously discussed) disappear (see panels 
$c$ and $d$ of Figure~\ref{Fig:Dl-dep}).
Since the width of each plateau is about five Doppler widths, it is 
important to observe that their merging, and therefore the disappearance of the 
effects of the interferences, starts when the intensity profiles of the two 
lines are still separated from each other (see panel $c$ of 
Figure~\ref{Fig:Dl-dep})\footnote{Note that the intensity profiles are in 
emission since we are considering the radiation scattered at 90$^{\circ}$ by 
an optically thin slab.}.

The effects of interferences become appreciable in the line-core when the 
separation between the two components is of the same order of magnitude as the 
width of the two interfering $J$-levels.
Indeed, in the limiting case in which the separation between the two levels 
goes to zero, the effects (and in particular the depolarizing effect) of the 
fine structure have to disappear, and the polarization pattern of a two-level 
atom with $J_u = L_u$ and $J_{\ell} = L_{\ell}$ must be recovered (because of 
the principle of spectroscopic stability).
This effect is shown in panels $e$ and $f$ of Figure~\ref{Fig:Dl-dep}, where 
a frequency separation between the two components of the same order of 
magnitude as the Einstein coefficient for spontaneous emission has been 
considered. 
It must be observed that the FS splitting of the various $J$-levels is always 
much larger (also in the case of hydrogen) than the natural width of the 
levels. 
For example, in the case of the Ly$\alpha$ line, the separation between the 
two FS components is of the order of 10$^{10}$ s$^{-1}$, while the Einstein 
coefficient for spontaneous emission is equal to 
\mbox{$6.265 \times 10^8$ s$^{-1}$}.
However, we have to remind that collisions (here neglected) may have an 
important role on these phenomena (both for their broadening and depolarizing
effect).

Another interesting result to be pointed out is that when the separation 
between the two components is smaller than the Doppler width of the two lines, 
so that the two plateaux completely merge, the FS depolarizing effect 
takes place on a wavelength interval of about five Doppler widths (the width of 
a single plateau), independently of the actual separation between the two 
components.

\subsection{The role of the continuum}
\label{Sect:continuum}
\begin{figure}[!t]
\centering
\includegraphics[width=\textwidth]{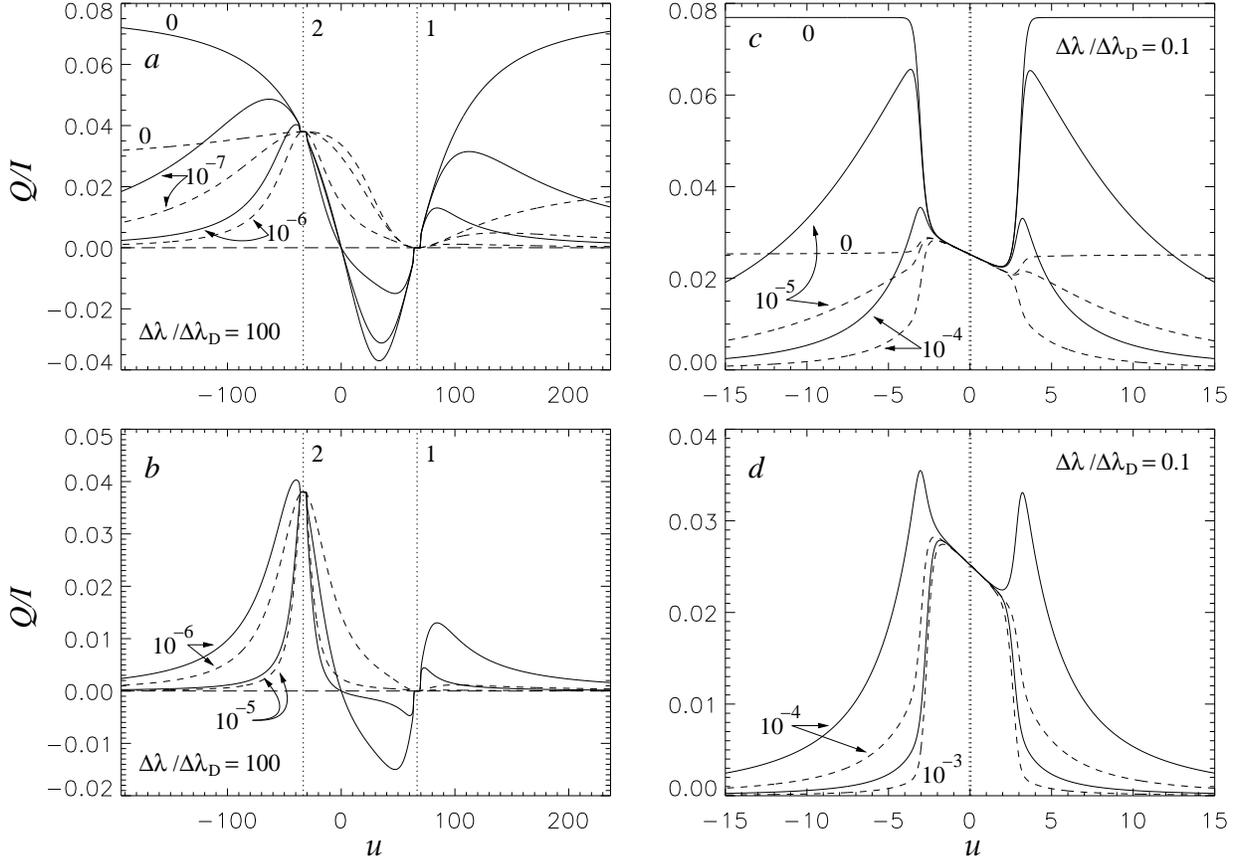}
\caption{\footnotesize{Fractional polarization profiles calculated according 
to Equation~(\ref{Eq:emer-pol}) for different values of the continuum 
$\varepsilon_I^{\, c}$, as obtained taking into account (solid line) and 
neglecting (dashed line) interferences between the two upper $J$-levels.
The values of $\varepsilon_I^{\, c}/\varepsilon_I^{\, \ell}({\rm max})$ 
corresponding to the various profiles are indicated in the plots.}}
\label{Fig:cont-dep}
\end{figure}

We now analyze how the fractional polarization profiles previously obtained 
taking into account only line processes are modified when the contribution 
of an unpolarized continuum is added according to Equation~(\ref{Eq:emer-pol}).
As shown by the various panels of Figure~\ref{Fig:cont-dep}, the main effect of 
the continuum is to make the fractional polarization vanish as one moves from 
the core of the lines toward the wings. 
Although this overall effect makes the patterns obtained taking into account 
and neglecting interferences coincide, if the continuum is not too strong, 
clear signatures of the interferences between different $J$-levels can still 
be noticed.

Let us consider first the case in which the wavelength separation between the 
two transitions is much larger than the Doppler width 
(Figure~\ref{Fig:cont-dep}, panels $a$ and $b$).
As far as the profile calculated taking into account interferences is 
concerned, we observe that if the continuum is not sufficiently strong, the 
maximum value of the polarization is not found at the center of transition 2, 
but at slightly shorter wavelengths. 
This circumstance can be seen in the observations of the Ca~{\sc ii} H and K 
system \citep[see][]{Ste80,Gan02}.
The amount of continuum needed in order to make the maximum signal falling at 
the center of transition 2 is the larger the smaller the wavelength separation 
between the two components.
The presence of the continuum reduces the negativity of the $Q/I$ pattern 
between the two transitions, it moves the negative minimum toward the 
wavelength position of transition 1, but it does not affect the wavelength 
position of the sign reversals.
Moreover, it produces a positive peak in the red wing of transition 1, so that 
an antisymmetrical profile across this component appears.
This latter effect, due to interferences between the different $J$-levels of 
the upper term, can be clearly observed across the Na~{\sc i} D$_1$ line
\citep[see][]{Ste97}.
As the value of the continuum is increased, the amplitude of this 
antisymmetrical pattern is decreased, and the overall profile becomes more 
similar to the one obtained neglecting interferences.
The amount of continuum needed for this signature to be cancelled out is 
the larger the smaller the wavelength separation between the two components
(this explains why this structure can be observed across the D$_1$ line
of Na~{\sc i}, but not across the corresponding Ba~{\sc ii} line at 
4934~{\AA}).

We consider now the opposite case in which the various components are very 
close to each other with respect to the Doppler width of the single lines 
(Figure~\ref{Fig:cont-dep}, panels $c$ and $d$). 
In this case, as previously discussed, interferences between different 
$J$-levels do not produce any observable effect across a spectral interval
of about five Doppler widths centered at the wavelength of the two transitions, 
irrespectively of the continuum value.
Just outside this interval, on the other hand, the $Q/I$ profile obtained 
taking into account interferences shows, if the continuum is not too strong, 
two peaks. 
These $Q/I$ peaks disappear as the value of the continuum is increased.
It should be emphasized that the physical origin of this two-peak structure, 
which is obtained only when interferences are taken into account, lies in 
the way the continuum is included in the slab model we are considering here.
This does not mean that it is an artifact, but it is clear that its reliability 
is intimately related to the suitability of the model that we are using.
We finally observe that for the same value of the continuum, the $Q/I$ 
profiles obtained taking into account interferences always show more extended 
wings than the corresponding profiles obtained neglecting them.

\subsection{The effect of radiative transfer}
All the results shown in this investigation are based on the slab model 
described in Section~\ref{Sect:model}, which allows us to study the 
polarization properties of the scattered radiation through the approximate 
analytical expression of Equation~(\ref{Eq:emer-pol}).
This allows us to analyze in great detail the atomic aspects of the problem, 
without introducing the complications due to radiative transfer processes in 
stratified model atmospheres\footnote{Note, however, that the second term in 
the right-hand side of Equation~(\ref{Eq:emer-pol}) describes the possibility 
of dichroism (i.e., selective absorption of polarization components), which is 
absent in optically thin media.}.

The analysis of the effects of radiative transfer on the polarization
signatures produced by interferences between different $J$-levels is beyond 
the scope of this paper. 
Nevertheless, we can make some brief, qualitative considerations on this topic.
The main consequence of radiative transfer is that, depending on its 
frequency across the spectral line profile, the scattered radiation comes from 
different atmospheric heights, characterized by different values of the 
anisotropy factor of the local radiation field.
The fractional polarization patterns previously obtained using a single value 
of the anisotropy factor will thus be modified by radiative transfer effects, 
through a modulation on the different values of $w$.
Such modifications will take place mainly between the center and the wings of 
the single spectral lines. 
Very likely, when radiative transfer effects in solar atmospheric models are 
considered, the amplitude and shape of the $Q/I$ profile around the line center
will be different from those shown in the plots of Figure~\ref{Fig:cont-dep}.
For example, we believe that the flat plateaux previously discussed will not 
be obtained, or will be different. 
On the other hand, the results of the analysis of the spectral intervals where 
the effects of interferences are negligible, and where the FS depolarization 
takes place, should remain valid, as well as the qualitative shapes of the 
signatures of interferences, as far as these appear in the far wings of the 
lines, where assuming a constant value of $w$ is quite a good approximation.

Finally, we emphasize that, despite of its simplicity, the slab model of 
Section~3 has already been applied with success for interpreting several 
peculiarities observed in scattering polarization signals 
\citep[e.g.,][]{Bel09}, thus showing that it represents a suitable 
approximation, at least for the modeling and understanding of the physical 
mechanisms producing polarization in the solar atmosphere.

\subsection{Application to particular $^2S-{^2P}$ multiplets}
\label{Sect:applications}
We consider now various $^2S-{^2P}$ multiplets of particular interest. 
If not explicitly stated, we keep assuming $w=0.1$ and $\bar{n}=10^{-3}$.
\begin{figure}[!t]
\centering
\includegraphics[width=\textwidth]{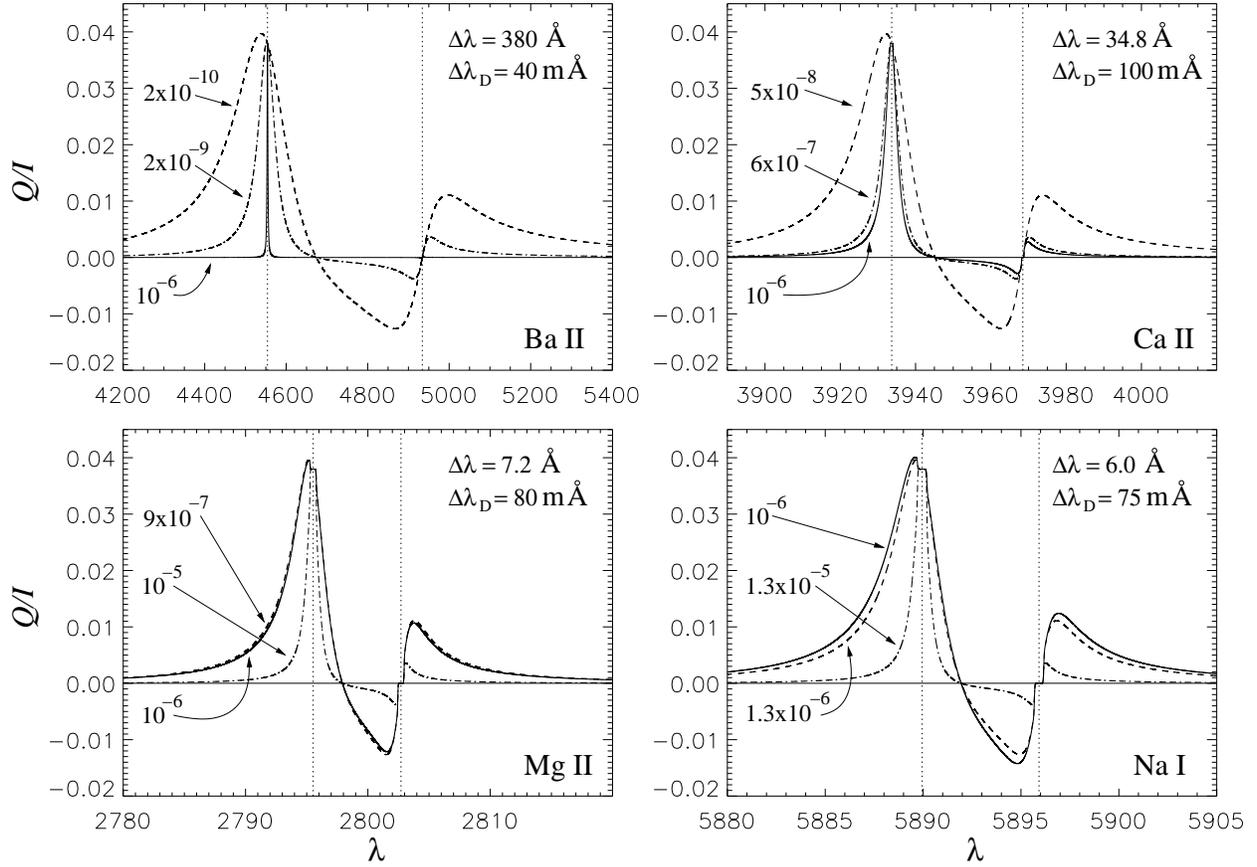}
\caption{\footnotesize{Fractional polarization patterns across the D$_1$ and 
D$_2$ lines of Ba~{\sc ii}, the H and K lines of Ca~{\sc ii}, the h and k 
lines of Mg~{\sc ii}, and the D$_1$ and D$_2$ lines of Na~{\sc i}, as 
calculated through Equation~(\ref{Eq:emer-pol}) for different values of the 
continuum, and taking into account interferences between the upper $J$-levels.
In all the panels, the profile plotted with solid line has been obtained 
assuming $\varepsilon_I^{\, c}/\varepsilon_I^{\, \ell}({\rm max})=10^{-6}$.
The profiles plotted with dashed and dash-dotted lines have been obtained
by choosing the continuum value for which the ratio between the amplitude of 
the negative minimum (in absolute value) and the value of $Q/I$ at line center 
of the K (or D$_2$) transition is equal to 1/3 and to 1/10, respectively.
The values of the continuum 
($\varepsilon_I^{\, c}/\varepsilon_I^{\, \ell}({\rm max})$) are indicated on 
the plots, as well as the wavelength separation between the two components 
($\Delta \lambda$), and the value of the Doppler width ($\Delta \lambda_D$).}}
\label{Fig:BaNa}
\end{figure}

Figure~\ref{Fig:BaNa} shows the fractional linear polarization patterns 
obtained across the D$_1$ and D$_2$ lines of Ba~{\sc ii}\footnote{It should be 
noticed that although these barium lines, being produced by a singly ionized 
atomic species, should be considered, from a spectroscopic point of view, of H 
and K type, they are often indicated as D$_1$- and D$_2$-type lines 
\citep[e.g.,][]{Ste97,Bel07} due to the similarities of the $Q/I$ scattering 
polarization signals that they produce at the limb with the corresponding 
signals produced by the D$_1$ and D$_2$ lines of Na~{\sc i}.}, the H and K 
lines of Ca~{\sc ii}, the h and k lines of Mg~{\sc ii}, and the D$_1$ and 
D$_2$ lines of Na~{\sc i}, taking into account the interferences between the 
two upper $J$-levels.
Although some of these elements (Ba, Mg and Na) have isotopes with non-zero 
nuclear spin, HFS has been neglected in this investigation.
For all these ions, the wavelength separation between the two lines of the 
multiplet is much larger than their Doppler width, so that, depending on the 
continuum emissivity, patterns similar to those shown in panels $a$ and $b$ 
of Figure~\ref{Fig:cont-dep} are obtained.

Comparing the profiles of the various ions plotted in Figure~\ref{Fig:BaNa}, it 
can be clearly observed how the amount of continuum needed for masking the 
signatures of interferences is the larger the smaller the separation 
between the two lines.
For example, in the case of barium, a continuum
$\varepsilon_I^{\,c}/\varepsilon_I^{\,\ell}({\rm max})=10^{-6}$ is already 
sufficient to completely cancel out the sign reversal between the two lines, 
and the ensuing antisymmetric pattern across the D$_1$ line.
These signatures, on the other hand, are still observable in calcium, 
magnesium and sodium, if the same continuum is considered.
From Figure~\ref{Fig:BaNa}, it can also be observed that for the wavelength 
separation of the magnesium (and sodium) lines, the amount of continuum needed 
in order to obtain a given ratio between the amplitude of the negative minimum 
(in absolute value) and the amplitude of the polarization peak of the k
(or D$_2$) line is more than 10 times stronger than in the case of the 
calcium lines.
In agreement with these considerations, the above-mentioned signatures of 
interferences between different $J$-levels can be clearly observed in calcium 
and sodium, but not in barium.

Unfortunately, there are no high sensitivity observations available as far as 
the h and k lines of Mg~{\sc ii} are concerned.
The fact that these two lines are very close to each other (their separation is 
similar to that between the sodium D-lines), and with very extended wings 
(so that, as in the case of calcium, the continuum is not completely reached 
between them), suggests that these signatures should be observable also across 
these lines. 
On the other hand, at the wavelength position of these lines the continuum 
is very strong, and has a very high degree of linear polarization, whose 
effect should be carefully taken into account.
We note that no sign reversal was found in the pioneering observations of 
these $Q/I$ signals performed by \citet{Hen87} using the Ultraviolet 
Spectrometer and Polarimeter on board of the {\it Solar Maximum Mission}
satellite, which however was not designed for detecting weak scattering 
polarization signals.

\begin{figure}[!t]
\centering
\includegraphics[width=0.5\textwidth]{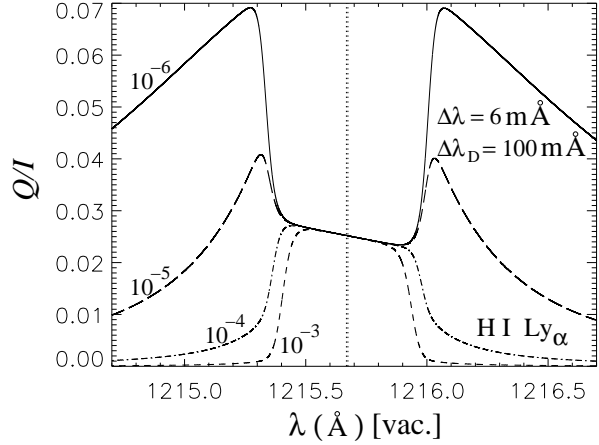}
\caption{\footnotesize{$Q/I$ profiles calculated according to 
Equation~(\ref{Eq:emer-pol}) in the Ly$\alpha$ line for different values of 
the continuum, taking into account interferences between the upper $J$-levels.
The values of the continuum 
($\varepsilon_I^{\, c}/\varepsilon_I^{\, \ell}({\rm max})$) are indicated in 
the figure, as well as the wavelength separation between the two components 
($\Delta \lambda$), and the Doppler width ($\Delta \lambda_D$).}}
\label{Fig:Ly-alpha}
\end{figure}

As a last example of $^2S-{^2P}$ multiplet, we consider the H~{\sc i} 
Ly${\alpha}$ line (see Figure~\ref{Fig:Ly-alpha}). 
Since the separation between the two components is in this case much smaller 
than the Doppler width of the line, the profiles that are obtained for 
different values of the continuum are very similar to those shown in panels
$c$ and $d$ of Figure~\ref{Fig:cont-dep}.
As previously pointed out, in this case interferences do not produce any 
effect in the core of the line, across a wavelength interval of about five 
Doppler widths. 
It can be noticed that this is the wavelength interval over which the 
depolarizing effect of fine structure takes place.
Signatures of interferences between the upper $J$-levels, consisting of two 
peaks appearing in the wings of the line, might be produced if the continuum 
is sufficiently weak (of the order of 10$^{-5}$ or smaller).
However, we already pointed out that these features in the calculated $Q/I$ 
profile might be due to the simple slab model that we are using 
and that might not be obtained once radiative transfer effects are properly 
taken into account.

The analysis carried out in this section on particular multiplets, 
characterized by different values of the wavelength separation between the 
two components, and of the Doppler widths, provides complete information on 
the role of interferences on $^2S-{^2P}$ doublets.
The extension of these results to other interesting doublets of this kind 
should not present particular difficulties.

\section{The $^5S-{^5P}$ multiplet}
In this section, we investigate the effects of interferences between different 
$J$-levels on the scattering polarization patterns produced across $^5S-{^5P}$ 
multiplets.
The possibility for the (common) lower level of the transitions of this 
multiplet to carry atomic polarization ($J_{\ell}=2$) allows us to investigate 
the effects of dichroism on the polarization signatures produced by 
interferences.
The analysis will be carried out on the Cr~{\sc i} triplet at 5207~{\AA} and 
on the O~{\sc i} triplet at 7773~{\AA}.

\subsection{The Cr~{\sc i} triplet at 5207~{\AA}}
\begin{figure}[!t]
\centering
\includegraphics[width=\textwidth]{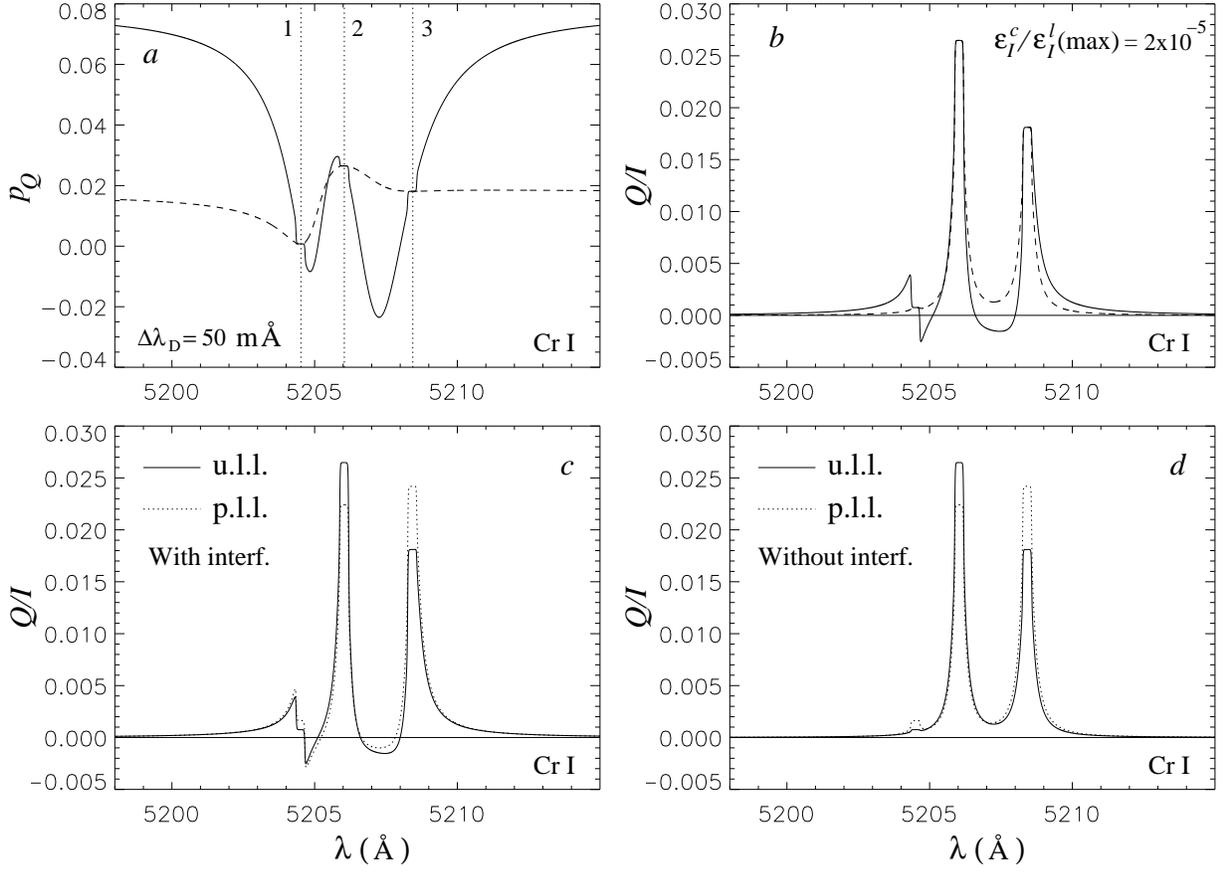}
\caption{\footnotesize{Panel $a$: 
$p_Q$ profiles obtained taking into account (solid line) and neglecting (dashed 
line) interferences between the upper $J$-levels. 
The vertical dotted lines indicate the wavelength positions of the various 
lines. Lower-level polarization has been neglected.
Panel $b$: same as panel $a$, but including the contribution of the continuum.
Panel $c$: $Q/I$ profiles obtained taking into account (dotted line) 
and neglecting (solid line) lower-level polarization. Interferences between 
different $J$-levels are taken into account. The same continuum as in panel
$b$ is considered. The profile obtained taking into account lower-level 
polarization has been calculated neglecting the second term in the 
right-hand side of Equation~(\ref{Eq:emer-pol}).
Panel $d$: same as panel $c$, but neglecting interferences between different 
$J$-levels. The abbreviations p.l.l. and u.l.l. stand for polarized and 
unpolarized lower level, respectively.}}
\label{Fig:Cr}
\end{figure}
The Cr~{\sc i} inverted\footnote{We recall that a multiplet is said to be 
regular when the energy of the $J$-levels increases with the value of $J$,
while it is said to be inverted in the opposite case.} triplet at 5207~{\AA} 
is composed by the following transitions: 
$J_{\ell}\!=\!2 \rightarrow J_u\!=\!1$ (line 1 at 5205.50~{\AA}),
$J_{\ell}\!=\!2 \rightarrow J_u\!=\!2$ (line 2 at 5206.04~{\AA}), and
$J_{\ell}\!=\!2 \rightarrow J_u\!=\!3$ (line 3 at 5208.42~{\AA}).
The HFS of the only stable isotope of chromium with non-zero 
nuclear spin ($^{53}$Cr, abundance 9.5\%) is neglected in the present 
investigation.
As in the previous sections, we consider the radiation scattered at 
90$^{\circ}$ by a slab of solar atmospheric plasma illuminated by the solar 
continuum radiation field ($w=0.1$, $\bar{n}=10^{-3}$).

As it can be observed in panel~$a$ of Figure~\ref{Fig:Cr}, the $p_Q$ profile 
obtained taking into account interferences between different $J$-levels shows 
in this multiplet two sign reversals, one between lines 1 and 2, and a larger 
one between lines 2 and 3.
The presence of a small plateau can be observed in the core of the three lines;
here, as in the case of the $^2S-{^2P}$ multiplet, interferences between 
different $J$-levels do not produce any observable effect.
Note that the profiles of panel $a$ of Figure~\ref{Fig:Cr} have been obtained 
neglecting lower-level polarization and stimulation effects: the value of 
$p_Q$ in the core of the lines is thus given by Equation~(\ref{Eq:pq-2Tcore}).

The effect of the continuum in masking the signatures of interferences is 
similar to the case of the $^2S-{^2P}$ multiplet. 
As shown in panel~$b$ of Figure~\ref{Fig:Cr}, the continuum makes the 
polarization vanish in the far blue wing of line 1 and in the far red wing
of line 3, it reduces the amplitude of the negative patterns between the lines, 
and it produces an antisymmetrical profile across transition 1 ($J_u=1$).
This last polarization signature, as well as a small ``sign reversal'' 
(with respect to the continuum polarization level) between lines 2 and 3, can 
be observed in Gandorfer's (2000) atlas of the second solar spectrum.

The lower level of this triplet, having $J_{\ell}=2$, can carry atomic 
polarization.
The first thing that has to be pointed out is the appreciable feedback that 
the presence of lower-level polarization has on the atomic polarization of the 
upper levels. 
As shown in panels~$c$ and $d$ of Figure~\ref{Fig:Cr}, the value of $Q/I$ 
(as calculated still neglecting the contribution of dichroism, given by the 
second term in the right-hand side of Equation~(\ref{Eq:emer-pol})) is modified 
by the presence of lower-level polarization in the core of the three lines 
(it is increased in lines 1 and 3, and decreased in line 2).
As far as the profiles obtained taking into account interferences between the 
upper $J$-levels are concerned (see panel $c$), we observe that lower-level 
polarization slightly modifies the negative pattern between lines 2 and 3, 
as well as the antisymmetrical profile across line 1.
Interferences between different $J$-levels do not produce any effect in the 
core of the lines also when lower-level polarization is taken into account.

\begin{figure}[!t]
\centering
\includegraphics[width=\textwidth]{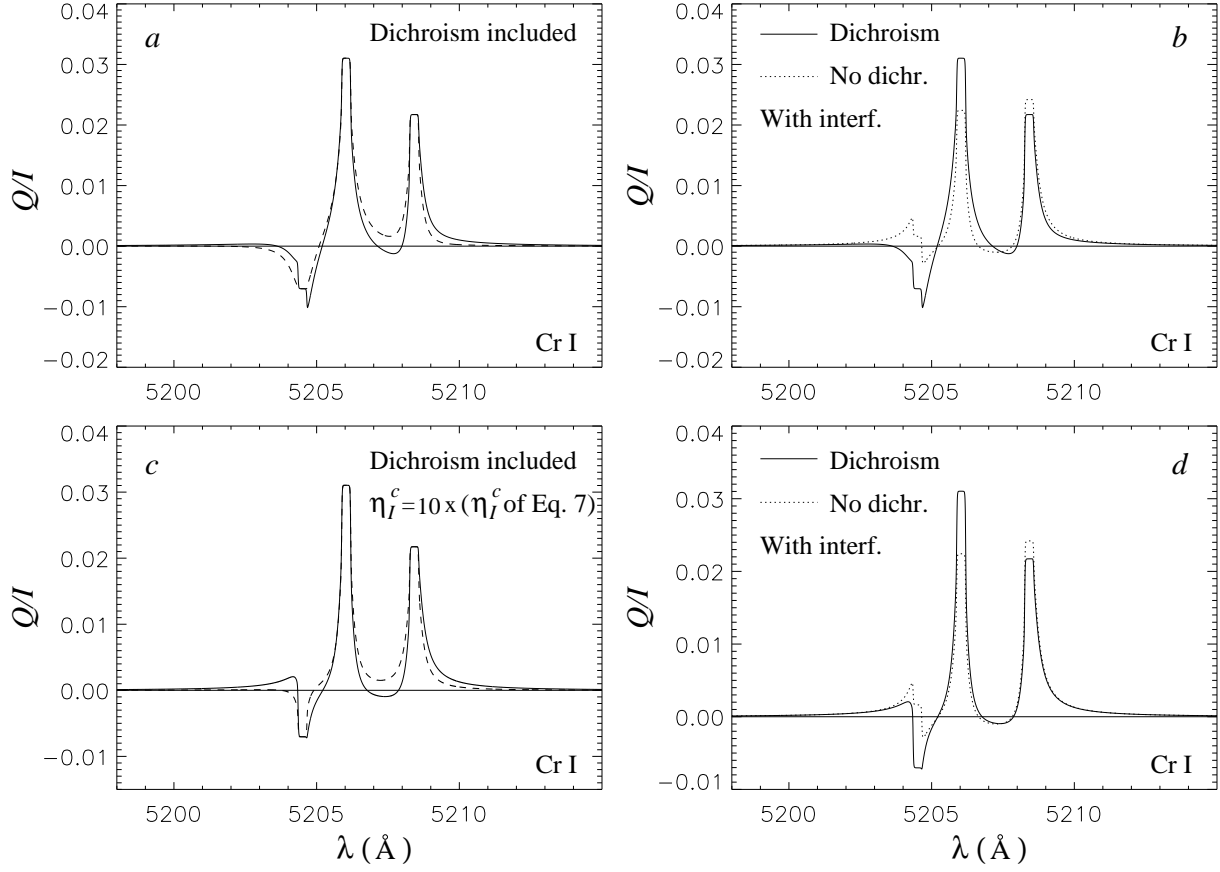}
\caption{\footnotesize{Panel $a$: $Q/I$ profiles calculated in the presence of 
dichroism (i.e., including the second term in the right-hand side of 
Equation~(\ref{Eq:emer-pol})), taking into account (solid line) and neglecting 
(dashed line) interferences between different $J$-levels. 
The value of $\varepsilon_I^{\, c}/\varepsilon_I^{\, \ell}({\rm max})$ is the 
same as in panel $b$ of Figure~\ref{Fig:Cr}, the value of $\eta_I^{\, c}$ has 
been calculated according to Equation~(\ref{Eq:etaIc}).
Panel $b$: $Q/I$ profiles calculated taking into account (solid line) and 
neglecting (dotted line) dichroism.
Interferences between different $J$-levels are taken into account.
The various parameters have the same values as in panel $a$.
Panels $c$ and $d$: same as panels $a$ and $b$, but assuming for 
$\eta_I^{\, c}$ a value 10 times larger than the one that is obtained through 
Equation~(\ref{Eq:etaIc}).}}
\label{Fig:Cr-dichr}
\end{figure}
When lower-level polarization is taken into account, $\eta_Q^{\, \ell}$ is 
non-zero, and also the second term in the right-hand side of 
Equation~(\ref{Eq:emer-pol}) (i.e., dichroism) brings a contribution to the 
polarization of the emergent radiation.
The effects of dichroism on the scattering polarization pattern of this 
multiplet can be clearly observed in Figure~\ref{Fig:Cr-dichr}.
As shown in panels $a$ and $b$ of this figure, the presence of dichroism 
modifies the value of $Q/I$ in the core of the three lines (note in particular 
that the polarization in the core of line 1 becomes negative), as well as the 
signatures of interferences between lines 2 and 3 (the wavelength positions of 
the sign reversals are changed), and across line 1 (the whole polarization 
pattern across this line is negative when dichroism is included, both 
when interferences are taken into account and when they are neglected).

In panels $a$ and $b$ of Figure~\ref{Fig:Cr-dichr}, the value of $\eta_I^{\,c}$ 
has been calculated according to Equation~(\ref{Eq:etaIc}), as explained in 
Section~\ref{Sect:model}.
However, if we consider this physical quantity as a free parameter, we observe 
that the signatures of interferences between different $J$-levels are
sensitive to its value.
This can be observed in panels $c$ and $d$ of Figure~\ref{Fig:Cr-dichr}.
If the value of $\eta_I^{\, c}$ is increased (with respect to the one 
calculated according to Equation~\ref{Eq:etaIc}), the dip produced by 
interferences in the red wing of line 1 (between lines 1 and 2), which is still 
observable in panels $a$ and $b$, gradually decreases until disappearing, while 
a positive bump appears in the blue wing of this line.
A profile with a sign reversal \citep[qualitatively similar to the one observed 
by][]{Gan00} is thus recovered in line 1. 
However, contrary to the profiles plotted in panel $c$ of Figure~\ref{Fig:Cr}, 
the negative minimum is now in the core of the line and not in its red wing. 
Indeed, the underlying physics is now the following: the positive peak in the 
blue wing is still due to interferences between different $J$-levels, while the 
negative minimum in the core of line 1 is produced by dichroism (see 
panel $d$ of Figure~\ref{Fig:Cr-dichr}).

\subsection{The O~{\sc i} triplet at 7773~{\AA}}
The O~{\sc i} triplet at 7773~{\AA} is composed by the following transitions
$J_{\ell}\!=\!2 \rightarrow J_u\!=\!1$ (line 1 at 7775.39~{\AA}),
$J_{\ell}\!=\!2 \rightarrow J_u\!=\!2$ (line 2 at 7774.16~{\AA}), and
$J_{\ell}\!=\!2 \rightarrow J_u\!=\!3$ (line 3 at 7771.94~{\AA}).
From an atomic point of view, this multiplet is very similar to the Cr~{\sc i} 
triplet at 5207~{\AA} previously investigated.
In particular, the FS splitting of the upper term, and therefore the wavelength 
separations among the various lines, is very similar in the two multiplets.
The only difference is that this is a regular triplet (i.e., not inverted, like 
the chromium one), so that the wavelength order of the transitions is the 
opposite with respect to the previous case.
\begin{figure}[!t]
\centering
\includegraphics[width=\textwidth]{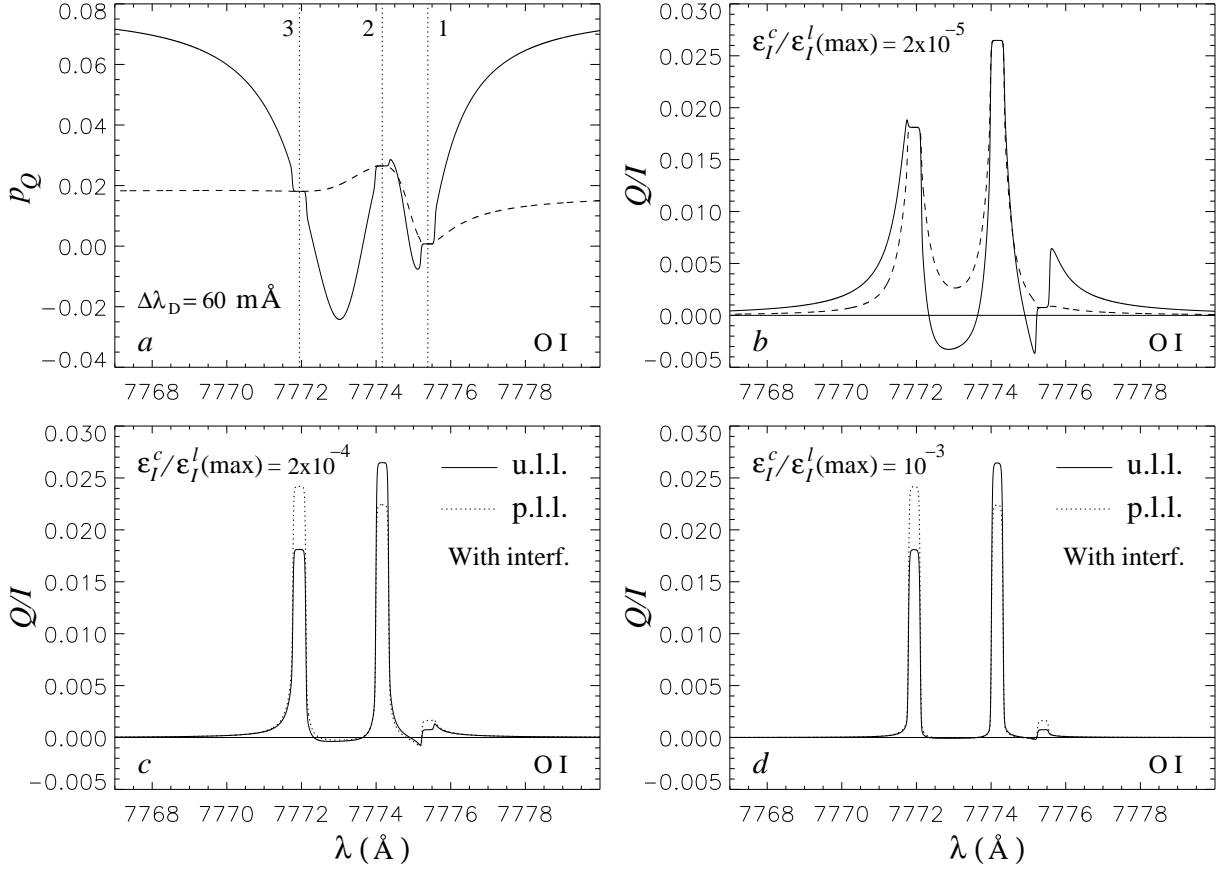}
\caption{\footnotesize{Panel $a$: $p_Q$ profiles obtained taking into account 
(solid line) and neglecting (dashed line) interferences between the upper 
$J$-levels. 
The vertical dotted lines indicate the wavelength positions of the various 
lines. Lower level polarization has been neglected.
Panel $b$: same as panel $a$, but including the contribution of the continuum.
Panel $c$: $Q/I$ profiles obtained taking into account (dotted line) and 
neglecting (solid line) lower-level polarization. 
Interferences between different $J$-levels are taken into account. 
The profile obtained taking into account lower-level polarization has been 
calculated neglecting the second term in the right-hand side of 
Equation~(\ref{Eq:emer-pol}).
The value of the continuum (higher than in panel $b$) is indicated on the plot.
Panel $d$: same as panel $c$, but for a higher value of the continuum. 
The abbreviations p.l.l. and u.l.l. stand for polarized and unpolarized lower 
level, respectively.}}
\label{Fig:O}
\end{figure}

Because of these similarities, not only the $p_Q$ patterns calculated taking 
into account and neglecting interferences are the same as in the Cr~{\sc i} 
triplet (compare panel $a$ of Figures~\ref{Fig:O} and \ref{Fig:Cr}) 
but also the $Q/I$ profiles obtained including the same amount of continuum 
are very similar in the two cases (compare panel $b$ of Figures~\ref{Fig:O} 
and \ref{Fig:Cr}).
The signatures of interferences between different $J$-levels are thus  
the same as in the Cr~{\sc i} triplet at 5207~{\AA}.

If a continuum $\varepsilon_I^{\, c}/\varepsilon_I^{\, \ell}({\rm max})$ 
stronger than in the case of chromium is considered (which seems to be a more 
suitable choice for these lines as they are much weaker than the corresponding 
chromium lines in the solar atmosphere), the signatures of interferences are, 
as expected, strongly reduced (see panels $c$ and $d$ of Figure~\ref{Fig:O}).

It should be noticed that the antisymmetric profile which is obtained 
across transition 1 when interferences between different $J$-levels and 
lower-level polarization are taken into account, and when the continuum is not 
too strong (see the dotted profile in panel $c$ of Figure~\ref{Fig:O}) is very 
similar to the one observed by \citet{Kel99} outside the solar limb (see their 
Figure 5).
On the other hand, the profiles plotted in panel $d$ of Figure~\ref{Fig:O}, 
obtained assuming a continuum sufficiently strong in order to cancel out
all the observational signatures of interferences, are in good qualitative 
agreement with the off-limb observation presented in figure 5 (top-right 
panel) of \citet{She03}.
This illustrates the sensitivity of the theoretical polarization profiles to 
the parameters of the model.

\begin{figure}[!t]
\centering
\includegraphics[width=\textwidth]{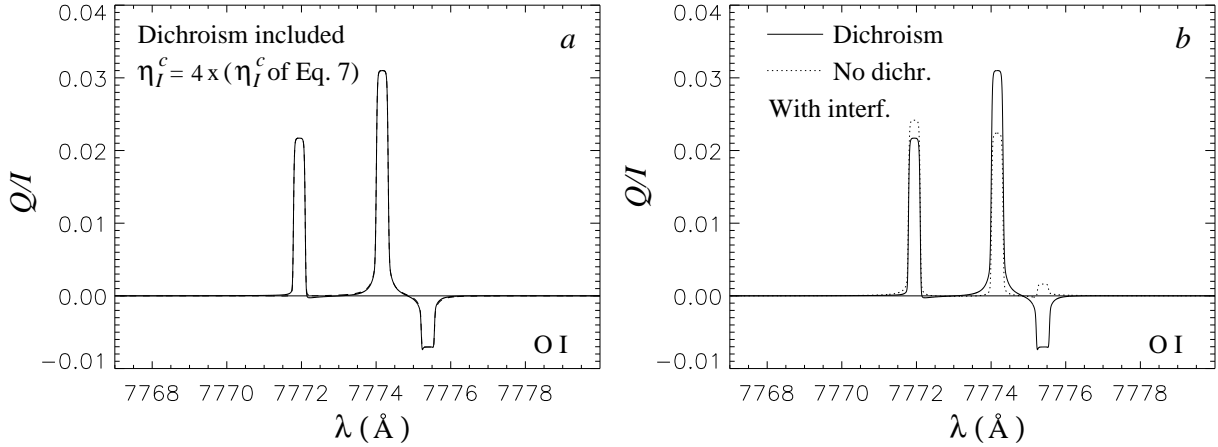}
\caption{\footnotesize{Panel $a$: 
$Q/I$ profiles calculated in the presence of dichroism, taking 
into account (solid line) and neglecting (dashed line) interferences between 
different $J$-levels (the two profiles cannot be distinguished). 
The value of $\varepsilon_I^{\, c}/\varepsilon_I^{\, \ell}({\rm max})$ is the 
same as in panel $d$ of Figure~\ref{Fig:O} ($10^{-3}$), the value of 
$\eta_I^{\, c}$ is four times larger than the one that is obtained through 
Equation~(\ref{Eq:etaIc}).
Panel $b$: 
$Q/I$ profiles calculated taking into account (solid line) and neglecting 
(dotted line) dichroism. 
Interferences between different $J$-levels are taken into account.
The various parameters have the same values as in panel $a$.}}
\label{Fig:O-dichr}
\end{figure}

Also the role of lower-level polarization is the same as in the case of 
chromium, both concerning its feedback on the atomic polarization of the upper
levels (see panels $c$ and $d$ of Figure~\ref{Fig:O}), as well as 
concerning the ensuing contribution of dichroism (see panels $a$ and 
$b$ of Figure~\ref{Fig:O-dichr}, and note the negative $Q/I$ signal present in 
the core of line 1).
If the same, large value of $\varepsilon_I^{\, c}$ as in panel $d$ of 
Figure~\ref{Fig:O} is considered, the signatures of interferences disappear 
also in the presence of dichroism (see panel $a$ of Figure~\ref{Fig:O-dichr}).
In particular, we note that although a large value of $\eta_I^{\, c}$ 
(larger than the one that would be obtained through Equation~(\ref{Eq:etaIc})) 
is considered, no positive bump is obtained in the red wing of line 1 (unlike 
the Cr~{\sc i} case shown in panel $c$ of Figure~\ref{Fig:Cr-dichr} which, 
being an inverted multiplet, shows such a positive bump in the blue wing of 
transition 1).
The profiles obtained taking into account dichroism are very similar to the 
ones calculated by \citet{JTB09} and show a very good agreement with the 
close to the limb, on-disk observation presented in Figure~4 of \citet{JTB09}.
Note that it is natural that off-limb and inside-limb observations may show 
different polarization features since the effects of dichroism are enhanced 
for on-disk observations.

\section{The $^3P-{^3S}$ multiplet: the role of lower-term interferences in the 
Mg~{\sc i} $b$-lines}
\label{Sect:MgI}
The multiplets considered in the previous sections allowed us to analyze the 
effects of interferences between different upper $J$-levels.
In order to investigate the role of interferences between different
$J$-levels of the lower term, we consider now the $^3P-{^3S}$ triplet of 
Mg~{\sc i} at 5178~{\AA}.
This triplet is composed by the following transitions: 
$J_{\ell}\!=\!2 \rightarrow J_u\!=\!1$ ($b_1$ line at 5183.60~{\AA}), 
$J_{\ell}\!=\!1 \rightarrow J_u\!=\!1$ ($b_2$ line at 5172.68~{\AA}), and 
$J_{\ell}\!=\!0 \rightarrow J_u\!=\!1$ ($b_4$ line at 5167.32~{\AA}).

We first note that under the hypothesis that the incident field is flat 
across the whole multiplet, if lower-term polarization is neglected, the 
emission coefficient $\varepsilon_Q^{\, \ell}$ is identically zero both for 
the two-term atom and for the corresponding four-level atom\footnote{As 
pointed out in LL04, this is due to the presence of the 6-$j$ symbol
\begin{equation}
	\Bigg\{ \!\!
	\begin{array}{ccc}
		1 & 1 & K \\
		L_u & L_u & L_{\ell} 
	\end{array}
	\!\! \Bigg\}
\end{equation}
in Equations~(\ref{Eq:eps2T}) and (\ref{Eq:WK-gen}), which is zero for $L_u=0$ 
and $K=2$.}.
We emphasize that this is a consequence of the flat-spectrum approximation 
required for the theory of the two-term atom to hold.
If we describe this triplet within the framework of a multi-level atom 
approach, so that we are allowed to consider a pumping field which varies 
among the various transitions, the $\varepsilon_Q^{\, \ell}$ coefficient will 
be in general different from zero, also under the hypothesis of unpolarized 
lower levels \citep[see][]{JTB09}.
Nevertheless, \citet{JTB99,JTB01} showed that the presence of a given amount of 
atomic polarization in the lower levels of this triplet is required 
in order to explain the observations presented in \citet{Ste00}, which show 
$Q/I$ signals of the same amplitude in all the three lines.
\begin{figure}[!t]
\centering
\includegraphics[width=\textwidth]{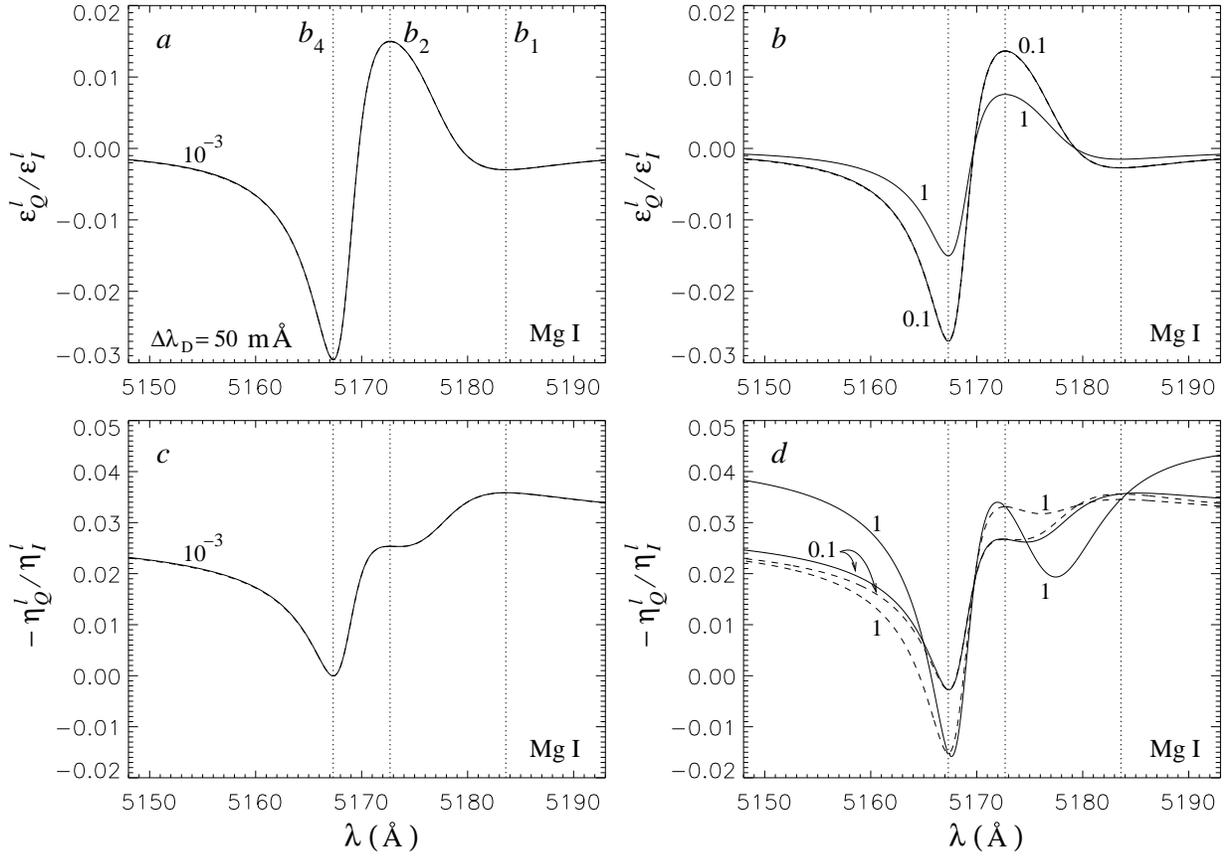}
\caption{\footnotesize{Panel $a$: 
$\varepsilon_Q^{\, \ell}/\varepsilon_I^{\, \ell}$ profiles calculated 
taking into account (solid line) and neglecting (dashed line) interferences 
between different lower $J$-levels, and assuming $\bar{n}=10^{-3}$.
Panel $b$: same as panel $a$ but for different values of $\bar{n}$ (indicated 
on the plot).
Panel $c$: same as panel $a$ but for the ratio 
$-\eta_Q^{\, \ell}/\eta_I^{\, \ell}$. 
Panel $d$: same as panel $c$ but for different values of $\bar{n}$ (indicated 
on the plot).
In panels $a$, $b$ and $c$ the profiles obtained taking into account and 
neglecting interferences cannot be distinguished.
The vertical dotted lines indicate the wavelength positions of the various 
lines.
Stimulated emission (not negligible when $\bar{n}$ assumes values of the 
order of 0.1 or larger) has been taken into account.}}
\label{Fig:MgI}
\end{figure}

Although lower-term polarization plays an important role in the Mg~{\sc i} 
$b$-lines, interferences between different lower $J$-levels are found to 
produce negligible effects on the 
$\varepsilon_Q^{\,\ell}/\varepsilon_I^{\,\ell}$ pattern of this 
multiplet (see panel $a$ of Figure~\ref{Fig:MgI}).
This result implies that the atomic polarization of the upper level is 
practically unaffected by the presence of this kind of interferences in 
the SEEs (as already pointed out in Section~10.21 of LL04).
Interestingly, interferences between different lower $J$-levels are found 
to play a negligible role also on the ratio $\eta_Q^{\,\ell}/\eta_I^{\, \ell}$
(see panel $c$ of Figure~\ref{Fig:MgI}).

Detailed analytical calculations performed in Section~10.21 of LL04 on the 
simpler $^2P-{^2S}$ multiplet show that interferences between different 
$J$-levels of the lower term are the smaller the larger the ratio $x_{\ell}$ 
between the FS splitting of the lower term and its natural width (given by 
$B(L_{\ell} \rightarrow L_u) \, J^0_0 = \bar{n} \, A(L_u \rightarrow L_{\ell}) 
\, (2L_u+1)/(2L_{\ell}+1)$).\footnote{When this ratio is very large, 
interferences between different lower $J$-levels vanish, and the levels become 
completely uncorrelated (like in a multi-level atom). 
Vice versa, when this ratio goes to zero, the lower $J$-levels are degenerate, 
and for the principle of spectroscopic stability the lower term has to behave 
like a $J$-level with $J_{\ell}=L_{\ell}$.}
Indeed, such a negligible effect of these interferences on the 
$\eta_Q^{\, \ell}/\eta_I^{\, \ell}$ profile is due to the low value of 
$\bar{n}$ that we are considering ($10^{-3}$).
As shown in panel $d$ of Figure~\ref{Fig:MgI}, if larger values of $\bar{n}$ (of
the order of 0.1 or larger) are considered, clear differences can be observed 
between the $\eta_Q^{\, \ell}/\eta_I^{\, \ell}$ profiles obtained taking into 
account and neglecting interferences between lower $J$-levels\footnote{Large 
$\bar{n}$ values are typical of masers.}.
Such differences are larger in the wings of the lines, while they 
disappear in the cores, where the effects of these interferences remain 
negligible also when large values of $\bar{n}$ are considered.
As discussed in LL04 for the $^2P-{^2S}$ multiplet, when lower-term 
polarization is taken into account, not only the interferences between 
different lower $J$-levels result to be sensitive to the value of $x_{\ell}$, 
but also the interferences between magnetic sublevels pertaining to the same 
$J$-level, as well as the populations of the various magnetic sublevels.
This explains the variation with $\bar{n}$ of the 
$\eta_Q^{\, \ell}/\eta_I^{\, \ell}$ profiles calculated neglecting the
interferences between lower $J$-levels.
It can be noticed that the sensitivity of these latter profiles to the 
value of $\bar{n}$ is limited to the cores of the lines (the asymptotic values
do not change significantly with $\bar{n}$), while a variation with $\bar{n}$ 
of the overall pattern (from the core to the far wings) takes place when 
interferences between different $J$-levels of the lower term are taken into 
account.

The effect of interferences between lower $J$-levels on the 
$\varepsilon_Q^{\, \ell}/\varepsilon_I^{\, \ell}$ profiles is found to be 
negligible also when large values of $\bar{n}$ are considered (see panel $b$ 
of Figure~\ref{Fig:MgI}).
Indeed, these profiles are much more sensitive to the value of $\bar{n}$ in the 
core of the lines than in the wings.
\begin{figure}[!t]
\centering
\includegraphics[width=\textwidth]{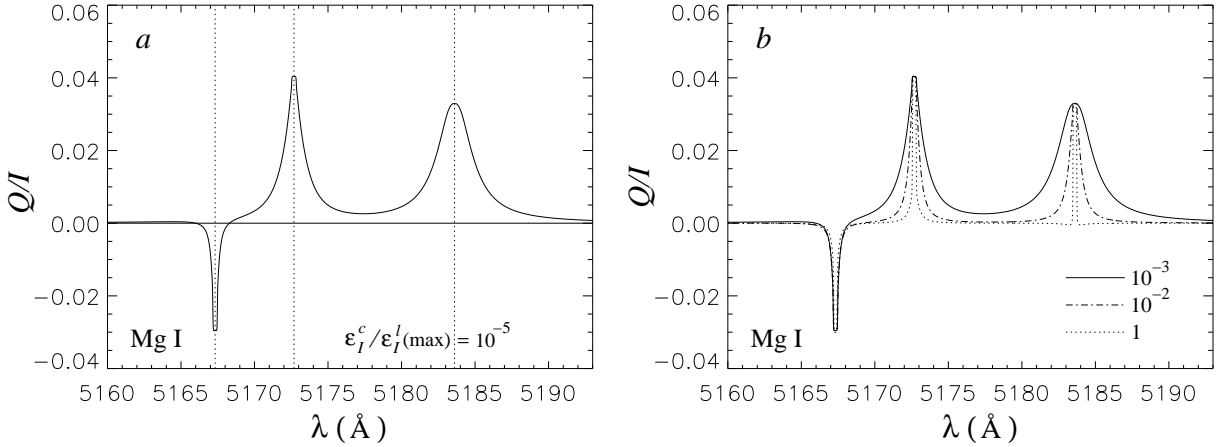}
\caption{\footnotesize{Panel $a$: $Q/I$ profiles calculated according to 
Equation~(\ref{Eq:emer-pol}) taking into account (solid line) and neglecting 
(dashed line) interferences between different lower $J$-levels, assuming 
$\bar{n}=10^{-3}$. 
The contribution of dichroism is included, the value of the continuum is 
indicated on the plot ($\eta_I^{\, c}$ is calculated according to 
Equation~(\ref{Eq:etaIc})). 
Note that the two profiles cannot be distinguished.
Panel $b$: $Q/I$ profiles calculated according to Equation~(\ref{Eq:emer-pol})
assuming different values of $\bar{n}$ (indicated on the plot).
The value of the continuum is the same as in panel $a$.
The profiles are calculated taking into account interferences between lower 
$J$-levels (note that for the values of the continuum here considered no 
differences can be observed between these profiles and the corresponding ones 
calculated neglecting such interferences).}}
\label{Fig:MgI-b}
\end{figure}

As expected from the previous discussion, interferences between lower 
$J$-levels do not produce any observable effect on the $Q/I$ profile calculated 
including the contribution of the continuum, taking into account dichroism, 
and assuming $\bar{n}=10^{-3}$ (see panel $a$ of Figure~\ref{Fig:MgI-b}).
Note that within the modeling assumptions here considered (and in particular 
the flat-spectrum approximation), a negative signal is obtained in the 
$b_4$ line (for understanding why the observations of \citet{Ste00} show 
positive signals in the three Mg~{\sc i} $b$-lines, see \citet{JTB09} 
and references therein).
The $Q/I$ profiles obtained for larger values of $\bar{n}$ are shown in 
panel~$b$ of Figure~\ref{Fig:MgI-b}. 
As it can be observed, the $Q/I$ pattern calculated through 
Equation~(\ref{Eq:emer-pol}) is quite sensitive to the value of $\bar{n}$ in 
the wings of the lines, while the line center polarization does not show any 
variation with $\bar{n}$.
We conclude pointing out that for the values of the continuum considered in 
Figure~\ref{Fig:MgI-b}, no differences can be observed between the $Q/I$ 
profiles obtained taking into account and neglecting interferences
between different lower $J$-levels, also when high values of $\bar{n}$ are 
assumed.

\section{The H$\alpha$ line}
\begin{figure}[!t]
\centering
\includegraphics[width=\textwidth]{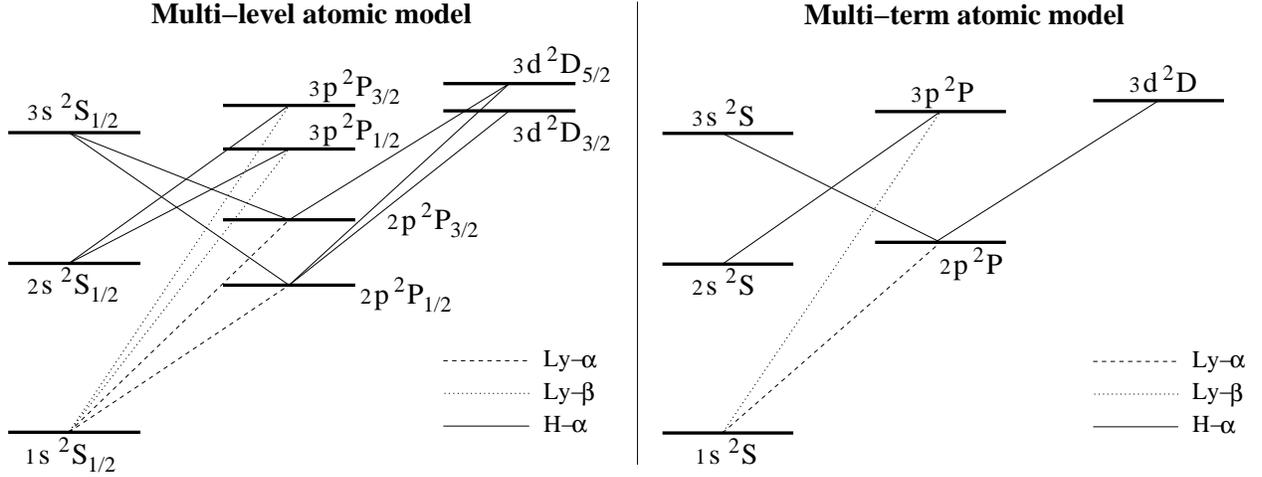}
\caption{\footnotesize{Grotrian diagrams of the multi-level (left) and 
multi-term (right) atomic models considered for the investigation of the
H$\alpha$ line. The separation among the various levels/terms is not on 
scale with the real energy differences present among them.}}
\label{Fig:Grotrian}
\end{figure}

Here we investigate the role of interferences between different $J$-levels 
on the scattering polarization profile of the H${\alpha}$ line.
Since this line is composed by seven FS components, belonging to three
different multiplets, a multi-term atomic model must be applied for the 
analysis of interferences in this line.
One of the lower terms is the upper term of the strong Ly$\alpha$ line, 
while one of the upper terms is also the upper term of the Ly$\beta$ line.
The inclusion of these lines in the atomic model is required for a correct 
analysis of the scattering polarization properties of H$\alpha$ 
\citep[e.g.,][]{Stp11}.
The Grotrian diagrams of the multi-level and multi-term atomic models 
considered are shown in Figure~\ref{Fig:Grotrian}.
We keep assuming $\bar{n}=10^{-3}$ and $w=0.1$ for all the lines considered in 
the model (the impact of interferences between different $J$-levels does not 
depend critically on the relative values of these quantities in the various 
lines).

\begin{figure}[!t]
\centering
\includegraphics[width=\textwidth]{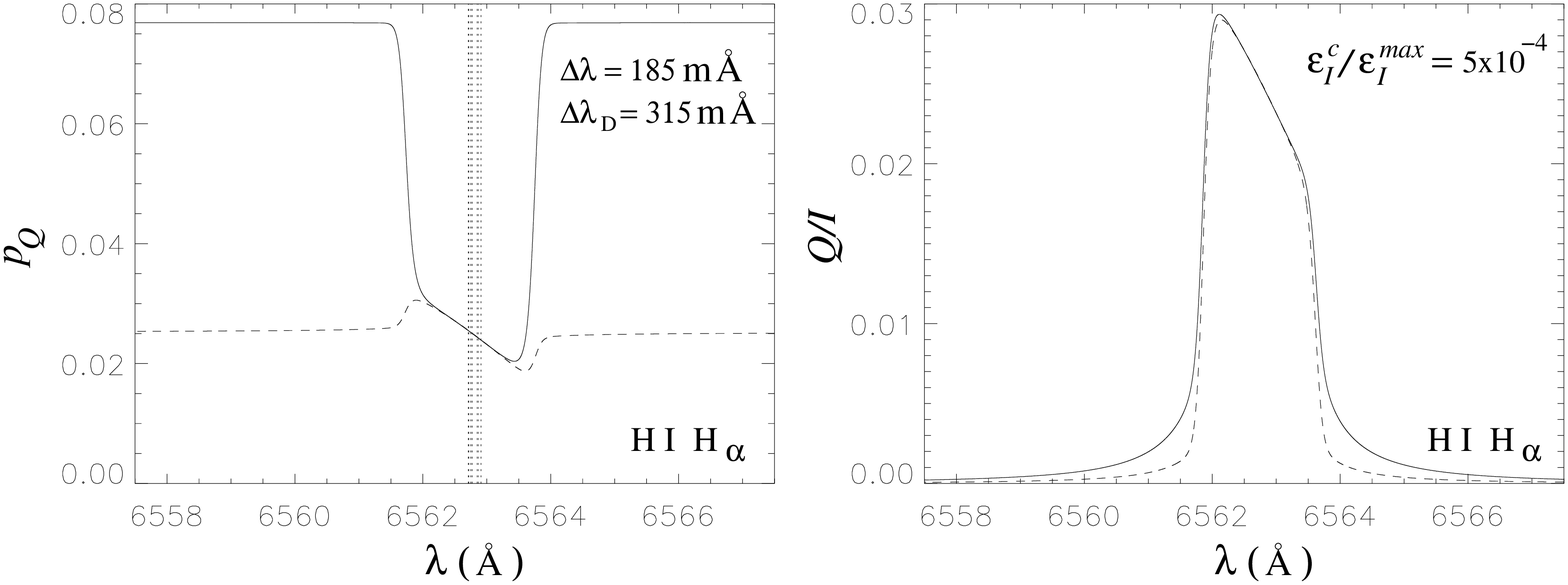}
\caption{\footnotesize{Left: $p_Q$ profiles calculated taking into account 
(solid line) and neglecting (dashed line) interferences between different 
$J$-levels. The vertical dotted lines indicate the wavelength position of the
seven components of this line. 
Right: $Q/I$ profile calculated according to Equation~\ref{Eq:emer-pol} 
(the contribution of the second term in the right-hand side has been neglected) 
taking into account (solid line) and neglecting (dashed line) interferences 
between different $J$-levels. The value of the continuum is indicated on the 
plot.
The profiles of both panels have been obtained taking into account 
lower-level/term polarization, whose influence is however negligible in solar
like atmospheres.}}
\label{Fig:H-alpha}
\end{figure}

Since the wavelength separation among the various components is much smaller 
than the Doppler width of the line, interferences between different $J$-levels 
do not produce any observable signature in the core of the line, across the 
whole wavelength interval over which the FS depolarization takes place (see 
left panel of Figure~\ref{Fig:H-alpha}).
In agreement with the previous results, the width of this wavelength interval 
is about 5 times the Doppler width of the line.

The polarization present in the $2p\,{^2P}$ lower term (the upper term of 
Ly$\alpha$) has been taken into account. 
However, its effect on the atomic polarization of the upper levels/terms 
through the SEEs is found to be negligible.

Also in this case, the $Q/I$ profile calculated taking into account 
interferences shows, for the same value of the continuum, slightly more 
extended wings than the profile calculated neglecting them (see right panel of 
Figure~\ref{Fig:H-alpha}). 
Finally, it must be pointed out that the line-core asymmetry shown by the $Q/I$ 
profile of the right panel of Figure~\ref{Fig:H-alpha} is not present when 
radiative transfer effects are fully taken into account in given semi-empirical 
models of the solar atmosphere \citep[see][]{Stp10}, although an asymmetric 
$Q/I$ profile similar to that observed by \citet{Gan00} can be produced in the 
presence of magnetic field gradients \citep[see][]{Stp10,Stp11}.

\section{The effects of a magnetic field}
We investigate now how the fractional polarization patterns described in the 
previous sections are modified by the presence of a magnetic field.
Before carrying out this analysis, we recall that in the multi-level atom 
approximation, any kind of correlation between different $J$-levels is
neglected by definition.
This implies that when this approach is applied in the presence of a magnetic 
field, it must be always assumed that the field is sufficiently weak for the 
Zeeman effect regime to hold (the splitting of the magnetic sublevels must be 
much smaller than the separation among the various $J$-levels).
This limitation is not required as far as the multi-term atom approximation is 
considered. 
In this latter case, magnetic fields going from the Zeeman effect regime to the 
complete Paschen-Back effect regime can be considered. 
As we will see (and as it is discussed in detail in LL04), in the incomplete 
Paschen-Back effect regime, interferences between different $J$-levels are 
at the origin of interesting phenomena (e.g., ``level-crossing'' and 
``anti-level-crossing'' effects), which may leave their signatures in the 
observed polarization profiles.

\subsection{The impact of the Hanle effect on the polarization pattern of the 
Mg~{\sc ii} h and k lines}
\label{Sect:MgII-mag}
We start considering the $^2S-\,^2P$ doublet of Mg~{\sc ii} at 2800~{\AA}.
The energy separation between the two $J$-levels of the upper term is 
sufficiently large for the Zeeman effect regime to hold, at least for the 
magnetic field intensities of the solar atmospheric plasma.
\begin{figure}[!t]
\centering
\includegraphics[width=\textwidth]{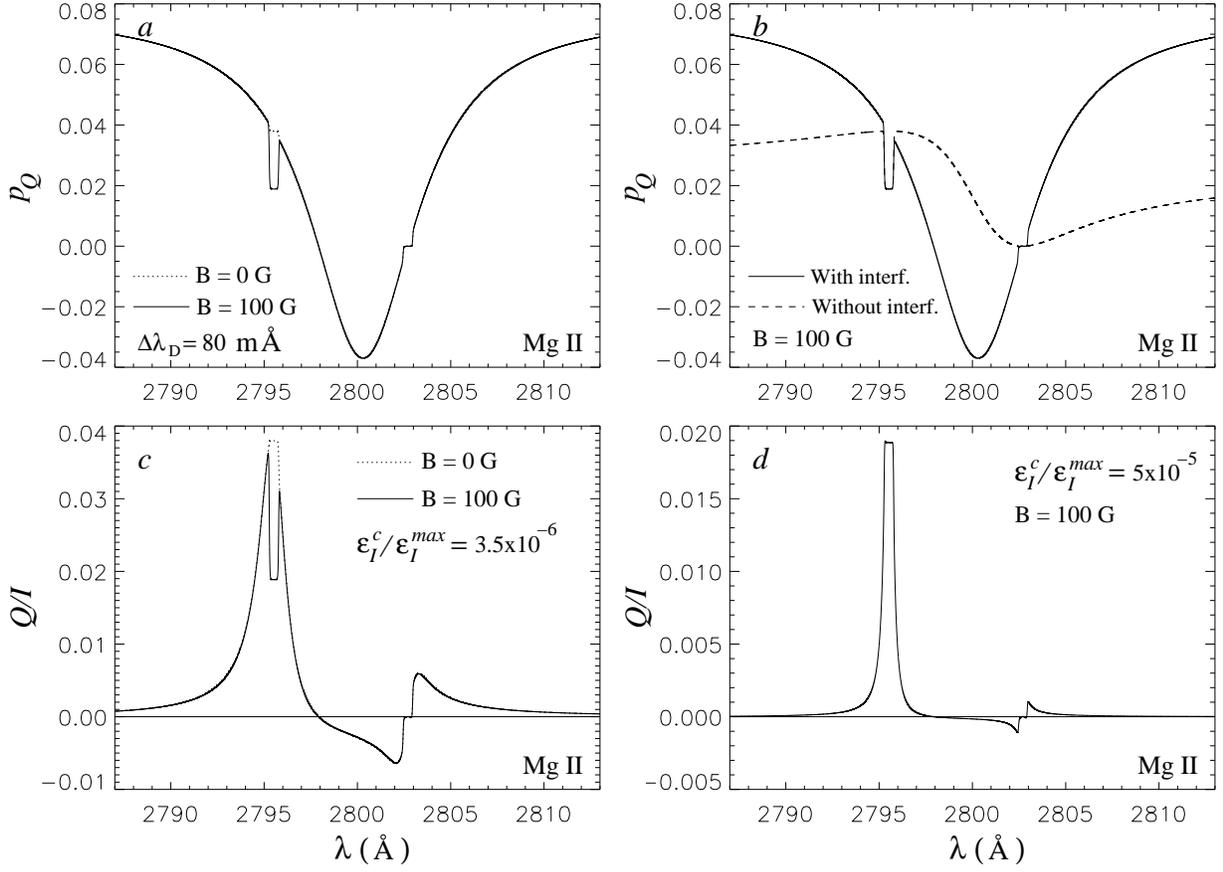}
\caption{\footnotesize{Effects of a horizontal magnetic field of 100~G 
perpendicular to the line of sight on the polarization pattern of the radiation 
scattered at 90$^{\circ}$ by an optically thin slab in the Mg~{\sc ii} doublet 
at 2800~{\AA}. 
Panel $a$: $p_Q$ profiles, calculated taking into account the interferences 
between the upper $J$-levels, in the absence (dotted line) and in the presence 
(solid line) of the above-mentioned magnetic field.
Panel $b$: $p_Q$ profiles calculated in the presence of the same field, taking 
into account (solid line) and neglecting (dashed line) interferences 
between the upper $J$-levels.
Panel $c$: same as panel $a$ but introducing the contribution of the continuum
according to Equation~(\ref{Eq:emer-pol}).
Panel $d$: same as panel $c$, but in the presence of a stronger continuum
(the profile corresponding to the unmagnetized case is not shown).}}
\label{Fig:Mg_mag}
\end{figure}

In the Zeeman effect regime, the magnetic sublevels pertaining to the same 
$J$-level split monotonically with the magnetic field strength. 
Because of the relaxation term proportional to $\nu_{mm^{\prime}}$ appearing 
in the SEEs (recall Equations~(\ref{Eq:SEE}) in Section~\ref{Sect:theory}, and 
the discussion therein), this splitting causes a decrease of the interferences 
between different magnetic sublevels pertaining to the same 
$J$-level\footnote{We observe that in a medium with cylindrical symmetry, these 
interferences are in general non-zero if the quantization axis is not 
taken along the symmetry axis (cf. the discussion in 
Section~\ref{Sect:model}).}, which in turn produces a decrease of the linear 
polarization degree of the emitted radiation with respect to the non magnetic 
case.
This is the Hanle effect for quantum interferences within the same $J$-level.
In the Mg~{\sc ii} doublet under investigation, this effect can be clearly 
observed in the core of the k line (see panel $a$ of Figure~\ref{Fig:Mg_mag}).

In this regime, on the other hand, the variation of the energy separation 
between magnetic sublevels pertaining to different $J$-levels is extremely 
small, so that interferences between different $J$-levels do not show 
significant variations with respect to the unmagnetized case.
The Hanle effect has therefore the same impact on the linear polarization 
patterns calculated taking into account and neglecting interferences, as 
shown in panel~$b$ of Figure~\ref{Fig:Mg_mag}.
As it can be clearly observed in the same figure, the Hanle effect takes 
place only in the core of the lines, right in the wavelength interval where the 
contribution of interferences between different $J$-levels is negligible.

Since the Hanle effect takes place only in the core of the lines, if the 
same continuum as in the unmagnetized case is considered, a two-peak $Q/I$ 
profile is obtained, the central dip being produced by the Hanle 
depolarization (see panel~$c$ of Figure~\ref{Fig:Mg_mag}).
This structure is gradually lost as the continuum intensity is increased.
On the other hand, if a continuum sufficiently strong to cancel out this 
two-peak structure is considered, also the signatures of interferences between 
different $J$-levels (such as the antisymmetric pattern across the h line) 
result to be strongly reduced (see panel $d$ of Figure~\ref{Fig:Mg_mag}).
As in Section~\ref{Sect:continuum}, we emphasize that the physical origin of 
this two-peak structure lies in the way the continuum is included in the slab 
model.

\begin{figure}[!t]
\centering
\includegraphics[width=\textwidth]{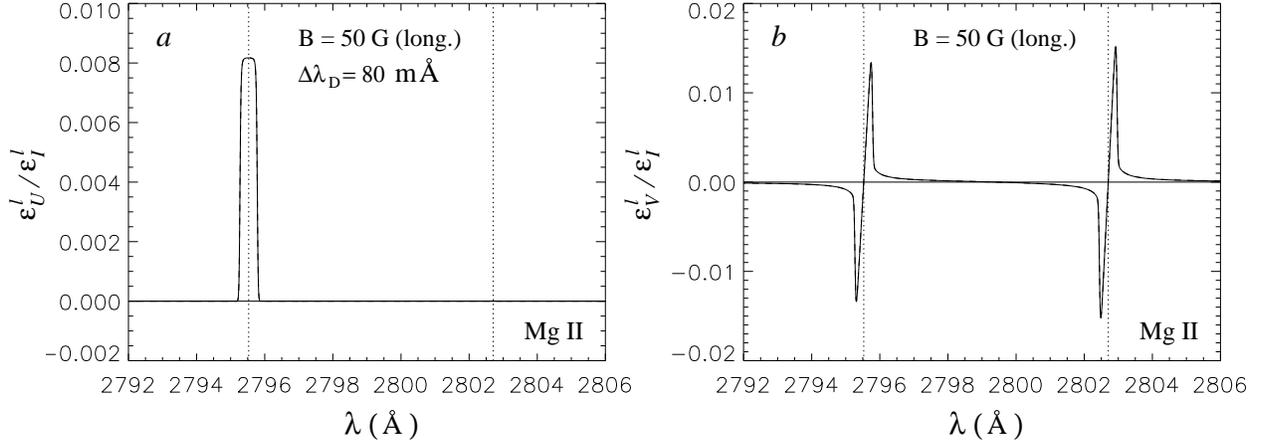}
\caption{\footnotesize{$\varepsilon_U^{\, \ell}/\varepsilon_I^{\, \ell}$ 
(panel $a$) and $\varepsilon_V^{\, \ell}/\varepsilon_I^{\, \ell}$ (panel $b$) 
profiles calculated taking into account (solid line) and neglecting (dashed 
line) interferences between different $J$-levels, in the presence of a 
longitudinal magnetic field of 50~G. The profiles obtained taking into account 
and neglecting interferences cannot be distinguished.}}
\label{Fig:Mg_mag_UV}
\end{figure}
In the presence of a magnetic field with a longitudinal component, an 
appreciable $\varepsilon_U^{\, \ell}/\varepsilon_I^{\, \ell}$ profile is 
obtained in the k line. 
This signal, due to the Hanle effect, appears in the core of the k line,
over the same wavelength interval where the Hanle depolarization of the 
$\varepsilon_Q^{\, \ell}/\varepsilon_I^{\, \ell}$ profile takes place.
This is also the wavelength interval (of about five Doppler widths) where 
interferences between different $J$-levels do not produce any appreciable 
effect on the $\varepsilon_Q^{\, \ell}/\varepsilon_I^{\, \ell}$ pattern. 
Indeed, the profiles obtained taking into account and neglecting such 
interferences cannot be distinguished (see panel $a$ of 
Figure~\ref{Fig:Mg_mag_UV}).
We note that a very weak antisymmetrical 
$\varepsilon_U^{\, \ell}/\varepsilon_I^{\, \ell}$ profile (of the order of
$10^{-8}$, not visible in the figure) is obtained in the core of the 
h line when interferences are taken into account.
This signal, too weak to be observable in practice, and of purely academic
importance, is probably due to the small variation of the interferences between 
different $J$-levels produced by the magnetic field in the Zeeman effect regime.
As previously pointed out, this kind of effect (i.e., Hanle effect for 
interferences between different $J$-levels) is generally negligible for 
magnetic fields in this regime, since the variation of the energy separation 
between the interfering magnetic sublevels (pertaining to different $J$-levels) 
is very small. 
These effects become more important as the Paschen-Back effect regime is 
reached.
In the presence of the same field, also a 
$\varepsilon_V^{\, \ell}/\varepsilon_I^{\, \ell}$ profile due to the Zeeman 
effect is obtained (see panel $b$ of Figure~\ref{Fig:Mg_mag_UV}).  
There are no differences between the 
$\varepsilon_V^{\, \ell}/\varepsilon_I^{\, \ell}$ profiles calculated taking 
into account and neglecting interferences.

The Mg~{\sc ii} h and k lines are an example of a doublet in which the 
separation between the two lines is much larger than their Doppler width. 
The results obtained in this section can be extended to any other
$^2S-{^2P}$ multiplet in which the two components are sufficiently separated 
from each other. The results can also be extended to different multiplets,
provided that the various lines are well separated, that lower-term 
polarization is absent or negligible (in this multiplet it is zero by 
definition), and that the same hypotheses here considered can be made.

\subsection{The effect of a magnetic field on the antisymmetric interference 
profile of the H/D$_1$-type lines}
In this section, we investigate the effects of a magnetic field on the 
antisymmetric interference pattern characterizing the emergent $Q/I$ profile 
across the $1/2-1/2$ transition of the $^2S-{^2P}$ multiplets.
No Hanle depolarization takes place in this line, since its line-core 
polarization is identically zero already in the absence of magnetic fields
(see, for example, the previous analysis of the Mg~{\sc ii} doublet at 
2800~{\AA}).
\begin{figure}[!t]
\centering
\includegraphics[width=\textwidth]{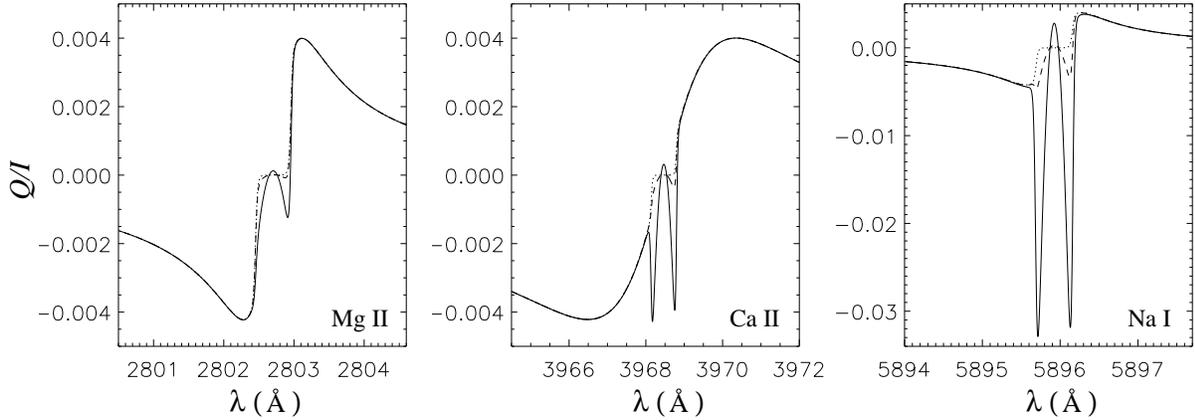}
\caption{\footnotesize{Modification of the antisymmetrical patterns produced 
by interferences between different $J$-levels in the emergent $Q/I$ profiles 
of the h line of Mg~{\sc ii} (left panel), of the H line of Ca~{\sc ii} (center 
panel), and of the D$_1$ line of Na~{\sc i} (right panel), due to the 
transverse Zeeman effect produced by a horizontal magnetic field perpendicular 
to the line of sight of 100~G (dashed line) and 300~G (solid line). 
The reference profiles corresponding to the unmagnetized case are shown by the 
dotted line.
The value of the continuum emissivity has been chosen in order to obtain the 
same amplitude of the antisymmetrical profile in the three lines.}}
\label{Fig:D1-mag}
\end{figure}
Nevertheless, it is interesting to observe how the interference pattern is 
modified by the presence of a magnetic field strong enough for the transverse 
Zeeman effect to be appreciable.
As shown in Figure~\ref{Fig:D1-mag}, when the typical signatures of the 
transverse Zeeman effect superimpose on the interference pattern, a peculiar 
profile, qualitatively similar to the one observed by \citet{Ste00} in the 
Na~{\sc i} D$_1$ line, is obtained (see also the theoretical investigations 
by Trujillo Bueno et al. 2002 and Casini \& Manso Sainz 2005, which took HFS 
into account).

The exact shape of the resulting profile depends on the relative amplitude and 
width of the antisymmetric interference pattern, and of the transverse 
Zeeman effect pattern (compare the three panels of Figure~\ref{Fig:D1-mag}).
The shape of the interference pattern is controlled by the continuum 
intensity (recall the discussion in Section~\ref{Sect:continuum}), while the 
shape of the Zeeman pattern by the magnetic field strength. 
As expected, the intensity of the magnetic field required for the Zeeman 
pattern to be appreciable on the overall profile is the higher the shorter the 
wavelength (compare the three panels of Figure~\ref{Fig:D1-mag}).

\subsection{The Ly$\alpha$ line case}
\begin{figure}[!t]
\centering
\includegraphics[width=0.5\textwidth]{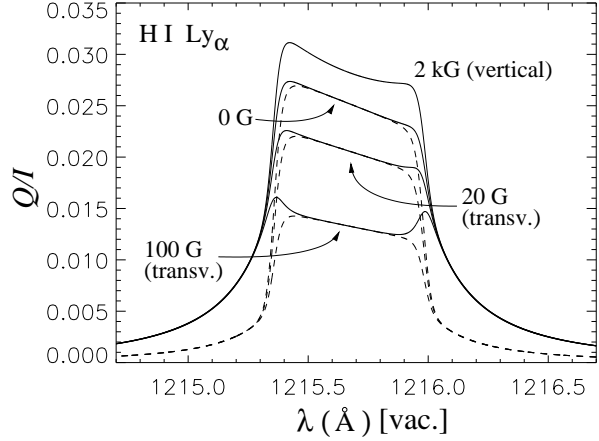}
\caption{\footnotesize{Fractional polarization profiles obtained taking into 
account (solid line) and neglecting (dashed line) interferences between the 
upper $J$-levels, in the presence of magnetic fields of different intensities 
and orientations (indicated on the plot). 
The profiles have been calculated according to Equation~(\ref{Eq:emer-pol}), 
assuming for the continuum the value
$\varepsilon_I^{\, c}/\varepsilon_I^{\, \ell}({\rm max})=6\times10^{-5}$.
Note that, as explained at the beginning of this section, for a magnetic field 
of 2~kG (i.e., in the incomplete Paschen-Back effect regime) only the two-term 
atom approach (with interferences between different $J$-levels) can be 
applied.}}
\label{Fig:Ly-alpha_mag}
\end{figure}

In order to investigate the effects of a magnetic field on a multiplet in which 
the separation among the various components is much smaller than their Doppler 
width, we consider now the Ly${\alpha}$ line.
As discussed in Section~\ref{Sect:applications}, interferences between 
different $J$-levels do not produce any observable signature in the core of 
this line.
Also in this case, the impact of the Hanle effect due to a magnetic field with 
strength in the Zeeman effect regime is the same on the polarization 
profiles obtained taking into account and neglecting interferences (see 
Figure~\ref{Fig:Ly-alpha_mag}).

In the Ly${\alpha}$ line, on the other hand, the separation between the two 
upper $J$-levels is sufficiently small for the incomplete Paschen-Back effect 
regime to be reached for magnetic fields of the order of 1~kG.
In this regime, the variation of the energy separation between pairs of 
magnetic sublevels pertaining to different $J$-levels is no longer negligible.
In particular, when the separation between two magnetic sublevels is of the  
same order of magnitude as their natural width (i.e., when the two sublevels 
overlap), the first term in the right-hand side of Equations~(\ref{Eq:SEE}) 
produces a significant modification of the corresponding coherence (generally 
an increase in absolute value), which may produce observable effects in the 
polarization of the emergent radiation.
It can be demonstrated that while pairs of magnetic sublevels with 
$\Delta M \ne 0$ can approach and cross each other (note that when this 
happens the relaxation term is exactly zero), pairs of magnetic 
sublevels with $\Delta M = 0$ can approach but never cross.
The terminology of `level-crossing' and `anti-level-crossing' 
\citep[see][]{Bom80} effects is often used to indicate the Hanle effect
produced by these particular behaviors of the magnetic sublevels in the 
incomplete Paschen-Back effect regime.
A detailed description of these phenomena can be found in LL04.

Here we want to focus the attention on the following aspect: the 
possibility, when interferences between different $J$-levels are considered, 
to have Hanle effect also in the presence of a ``vertical'' magnetic field.
As discussed in Section~\ref{Sect:model}, in a medium with cylindrical symmetry 
around a given direction (e.g., the vertical), if the quantization axis is 
taken along this direction, it can be shown that no quantum interferences 
can be induced between pairs of magnetic sublevels pertaining to the 
same $J$-level.
Such interferences, which are the only ones to be accounted for in a 
multi-level atom, also remain zero in the presence of a magnetic field 
directed along the symmetry axis of the incident radiation, since this does 
not break the symmetry of the problem. 
Since the magnetic field does not modify the populations (note that under the 
same hypotheses population imbalances can be induced among the various magnetic 
sublevels), it follows that, as known, for the multi-level atom model there is 
no Hanle effect in the presence of a vertical magnetic field.

This is no more true when interferences between different $J$-levels are 
considered.
As previously shown, in a medium with cylindrical symmetry, interferences 
between magnetic sublevels with the same value of $M$, pertaining to different 
$J$-levels, are actually non-zero.
These interferences are modified by a vertical magnetic field, in the 
incomplete Paschen-Back effect regime, with appreciable effects on the 
polarization of the scattered radiation.
This particular example of Hanle effect for interferences between different 
$J$-levels can be clearly observed in Figure~\ref{Fig:Ly-alpha_mag}: in the 
presence of a vertical magnetic field of about 2~kG (thus in the incomplete 
Paschen-Back effect regime), the amplitude of the scattering polarization 
signal is increased with respect to the zero-field case.
This enhancement of the polarization due to a vertical magnetic field was 
already pointed out by \citet{JTB02} for the Na~{\sc i} D$_2$ line, and by 
\citet{Bel07} for the Ba~{\sc ii} D$_2$ line. However, it should be observed 
that in these latter cases the effect is due to interferences between different 
HFS $F$-levels, and not between different $J$-levels as in the Ly${\alpha}$ 
line (the physical mechanism is exactly the same).
It is clear that this effect is of more practical interest when HFS is present, 
since magnetic fields relatively weak are sufficient for the incomplete 
Paschen-Back effect regime of HFS to be reached.

\subsection{Hanle effect and lower-level polarization}
\begin{figure}[!t]
\centering
\includegraphics[width=\textwidth]{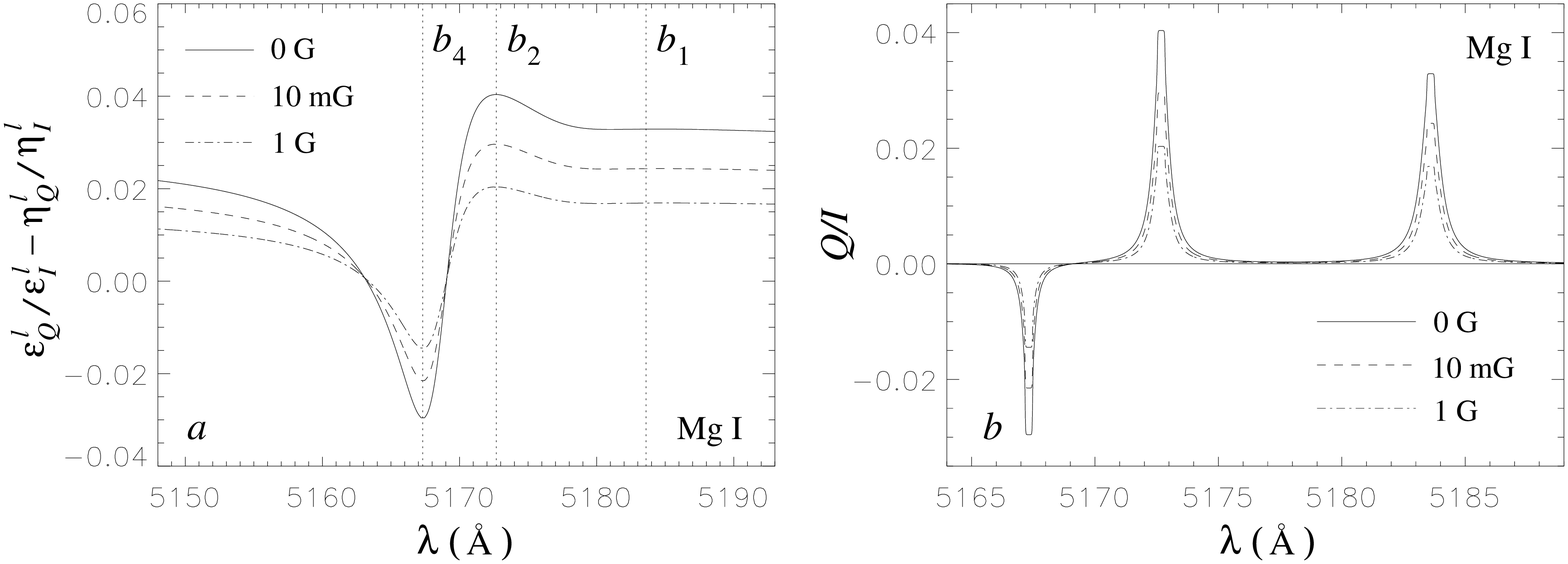}
\caption{\footnotesize{Fractional polarization profiles obtained in the 
presence of a horizontal magnetic field perpendicular to the line of sight of 
various intensities (see plots), neglecting (panel $a$) and taking into 
account (panel $b$) the contribution of the continuum. 
The value of the continuum used to calculate the profiles of panel $b$ is the 
same as in panel $a$ of Figure~\ref{Fig:MgI-b}.
All the profiles have been obtained taking into account interferences 
between lower $J$-levels, and assuming $w=0.1$ and $\bar{n}=10^{-2}$.}}
\label{Fig:wing-HE}
\end{figure}

We analyze here the effect of a magnetic field on the Mg~{\sc i} triplet at 
5178~{\AA}.
As discussed in Section~\ref{Sect:MgI}, the polarization of the radiation 
emitted in these lines is very sensitive to the atomic polarization of the 
lower levels.
The so-called lower-level Hanle effect can thus be clearly observed in this 
triplet.
If the lower level is metastable (as in this case), the lower-level Hanle 
effect is sensitive to rather weak magnetic fields.
In these lines, assuming $\bar{n}=10^{-2}$, it can already be appreciated for 
magnetic field intensities of the order of a few mG (see 
Figure~\ref{Fig:wing-HE}), while for a magnetic field of 1~G it is already 
saturated.

Also in the presence of magnetic fields, interferences between different lower 
$J$-levels do not produce any observable signature on the $Q/I$ profile of the 
emergent radiation, and they can be safely neglected.

An interesting property of the lower-level Hanle effect is that it does not 
vanish in the wings of the lines, but it depolarizes the whole pattern
(see panel $a$ of Figure~\ref{Fig:wing-HE}).
As a consequence, if the same continuum as in the unmagnetized case is 
considered, no double-peak structures are obtained as a result of the Hanle
depolarization (see panel $b$ of Figure~\ref{Fig:wing-HE}).

We recall that the property of the (upper-level) Hanle effect to vanish in the 
wings of the lines is strictly verified under the assumptions of unpolarized 
lower level, no stimulation effects, and no elastic collision (see Section~10.4 
of LL04 for an analytical proof).
It is clear that this property remains valid whenever stimulation effects are 
very weak (like in the solar atmosphere), and whenever the influence of 
lower-level polarization and collisions on the polarization properties of the 
line is negligible.
An analysis of the influence of elastic collisions on the Hanle effect can be 
found in Section~10.6 of LL04.
A detailed analysis of the physical conditions under which 
``wing-Hanle-effect'' can be observed will be presented in a forthcoming paper.

\section{Conclusions}
We have investigated the effects of quantum interferences between different 
$J$-levels on the linear polarization pattern of the radiation scattered at 
90$^{\circ}$ by a slab of stellar atmospheric plasma.
The investigation has been carried out within the framework of the quantum
theory of polarization presented in LL04, which is based on the flat-spectrum 
approximation.

We started focusing our attention on the $^2S-{^2P}$ doublet.
We analyzed the effects of the interferences between the two upper
$J$-levels as a function of the wavelength separation ($\Delta \lambda$) 
between the two FS components, and assuming a finite Doppler width 
($\Delta \lambda_D$) for the two spectral lines, in the absence of the 
continuum.
The most important results of this part of our study, carried out for 
the unmagnetized case, and neglecting the effects of collisions and stimulated 
emission, are the following.
\begin{enumerate}
	\item{The effects of interferences between different $J$-levels are 
		negligible in the core of the two lines, while they become 
		important moving from the center to the wings (see panel $a$ 
		of Figure~\ref{Fig:pq-gen}).
		If the separation $\Delta \lambda$ between the two lines is 
		much larger than their Doppler width, the shape of the overall 
		interference pattern (shown in panel $a$ of 
		Figure~\ref{Fig:pq-gen}) does not depend on the particular 
		value of $\Delta \lambda$.}
	\item{In the core of the two lines, the	fractional polarization 
		profiles obtained taking into account and neglecting such 
		interferences show flat ``plateaux'' of about 
		$5 \Delta \lambda_D$, across which they perfectly coincide
		(see panels $b$ and $c$ of Figure~\ref{Fig:pq-gen}).}
	\item{When $\Delta \lambda \lesssim 5 \Delta \lambda_D$ (so that the 
		two plateaux merge), all the signatures due to interferences 
		disappear between the two-lines (note that this merging starts 
		when the intensity profiles are still well separated from each 
		other; see panel $c$ of Figure~\ref{Fig:Dl-dep}).}
	\item{When $\Delta \lambda < \Delta \lambda_D$ (so that the two
		plateaux completely merge), the fine structure depolarization 
		takes place on a wavelength interval of about 
		$5 \Delta \lambda_D$ (the width of a single plateau), 
		irrespectively of the actual separation between the two 
		components (see panels $d$, $e$, and $f$ of 
		Figure~\ref{Fig:Dl-dep}).}
	\item{Signatures of interferences become appreciable in the 
		line-core when the separation between the interfering $J$-levels
		is of the same order of magnitude as their natural width 
		(see panels $e$ and $f$ of Figure~\ref{Fig:Dl-dep}; note that 
		this circumstance is not met by any real multiplet).}
\end{enumerate}

Although we believe that the flat plateaux that appear in the core of the lines 
may not be obtained once radiative transfer effects in realistic solar model 
atmospheres are taken into account, the fact that across the corresponding 
wavelength intervals the effects of interferences are negligible is a physical 
result that is expected to remain valid also when full radiative transfer is 
properly considered. 
A detailed analysis of the physical and mathematical origin of these intervals, 
and of their width (of the order of $5 \Delta \lambda_D$), has thus been 
carried out (see the Appendix).

We then analyzed how the signatures of interferences are masked when 
the contribution of an unpolarized continuum is taken into account, finding
the following main results.
\begin{enumerate}
	\item{The amount of the continuum emissivity needed to mask the 
		signatures of interferences is the larger the smaller the 
		wavelength separation between the interfering lines (see 
		Figures~\ref{Fig:cont-dep} and \ref{Fig:BaNa}).}
	\item{In the case of the $^2S-{^2P}$ multiplet, the effect of the 
		continuum is to reduce the amplitude of the negative minimum 
		between the two lines, to move it toward the $1/2 - 1/2$ 
		transition, and to produce an antisymmetrical pattern across 
		this transition (see Figure~\ref{Fig:cont-dep}).
		For a given value of the continuum emissivity, the amplitude 
		of the above-mentioned antisymmetrical pattern is the smaller 
		the larger the separation between the two components of the 
		multiplet (see Figure~\ref{Fig:BaNa}).}
\end{enumerate}
The above-mentioned results are not all limited to the $^2S-{^2P}$ doublet, 
but they can be generalized to any other multiplet. 

We carried out an analysis of the signatures produced by interferences between 
different $J$-levels on the following multiplets:
\begin{itemize}
	\item{Ba~{\sc ii} $^2S-{^2P}$ doublet (4554~{\AA} and 4934~{\AA} 
		resonance lines);}
	\item{Ca~{\sc ii} $^2S-{^2P}$ doublet (H and K lines);}
	\item{Mg~{\sc ii} $^2S-{^2P}$ doublet (h and k lines);}
	\item{Na~{\sc i} $^2S-{^2P}$ doublet (D$_1$ and D$_2$ lines);}
	\item{H~{\sc i} $^2S-{^2P}$ doublet (Ly$\alpha$);}
	\item{Cr~{\sc i} $^5S-{^5P}$ triplet at 5207~{\AA};}
	\item{O~{\sc i} $^5S-{^5P}$ triplet at 7773~{\AA};}
	\item{Mg~{\sc i} $^3P-{^3S}$ triplet (b$_1$, b$_2$ and b$_4$ lines);}
	\item{H~{\sc i} H$\alpha$ (line composed of seven FS 
		components belonging to three different multiplets).}
\end{itemize}

The analysis of the Cr~{\sc i} and O~{\sc i} triplets allowed us to investigate 
in some details the combined effect on the emergent scattering polarization 
profiles of interferences between different $J$-levels, lower-level 
polarization (see Figures~\ref{Fig:Cr} and \ref{Fig:O}), and dichroism (see 
Figures~\ref{Fig:Cr-dichr} and \ref{Fig:O-dichr}), for different values of the 
background continuum.

The analysis of the Mg~{\sc i} b-lines allowed us to investigate the role 
of interferences between different $J$-levels of the lower term.
We found that for the typical solar values of $\bar{n}$ (the mean number of 
photons per mode of the pumping radiation field), their effect on the 
scattering polarization profiles is completely negligible (see panels $a$ and 
$c$ of Figure~\ref{Fig:MgI}, and panel $a$ of Figure~\ref{Fig:MgI-b}).
Values of $\bar{n}$ of the order of 0.1 or larger are needed in order to 
observe their signatures on the ratio $\eta_Q/\eta_I$, their effect being in 
any case negligible in the core of the lines (see panel $d$ of 
Figure~\ref{Fig:MgI}).
The atomic polarization of the upper levels is practically unaffected by 
the presence of these interferences in the SEEs,
also for high values of $\bar{n}$ (see panel $b$ of Figure~\ref{Fig:MgI}).
The signatures of interferences between different lower $J$-levels (appreciable 
only when high values of $\bar{n}$ are considered) are in any case strongly 
masked by the presence of the continuum.
Our calculations also showed an appreciable sensitivity of the $Q/I$ profile 
of the emergent radiation to the value of $\bar{n}$ (see panel $b$ of 
Figure~\ref{Fig:MgI-b}).

We finally investigated whether or not, and to which extent, the signatures 
due to interferences between different $J$-levels are modified by the presence
of a magnetic field. The results can be summarized as follows.
\begin{enumerate}
	\item{In the Zeeman effect regime, the influence of the magnetic 
		field on the values of the interferences between different 
		$J$-levels is extremely small. 
		The Hanle effect thus leaves the same signatures on the 
		fractional polarization patterns calculated taking into 
		account and neglecting interferences between different 
		$J$-levels. 
		As shown by the profiles calculated for the Mg~{\sc ii} h and 
		k lines, as far as the
		$\varepsilon_Q^{\, \ell}/\varepsilon_I^{\, \ell}$ 
		pattern is concerned, the Hanle effect takes place in the 
		core of the k line, right in the wavelength interval 
		where the effect of interferences is negligible (see 
		Figure~\ref{Fig:Mg_mag}).
		Also the $\varepsilon_U^{\, \ell}/\varepsilon_I^{\, \ell}$ 
		signal, produced by the Hanle effect in the same line in 
		the presence of a longitudinal field, appears across this 
		wavelength interval (see panel $a$ of 
		Figure~\ref{Fig:Mg_mag_UV}).}
	\item{In the incomplete Paschen-Back effect regime, interferences
		between different $J$-levels are significantly modified by 
		the magnetic field, thus producing observable effects on the 
		emergent radiation.
		The physical mechanisms at the origin of these effects
		(e.g., level-crossings, anti-level-crossings, 
		alignment-to-orientation conversion mechanism) are described 
		in detail in LL04. 
		In this paper, we focused our attention on the possibility
		to have Hanle effect also in the presence of ``vertical'' 
		fields when interferences between different $J$-levels are 
		taken into account (see Figure~\ref{Fig:Ly-alpha_mag}).
		We should remember that the Paschen-Back effect regime is more 
		typical for HFS multiplets.
		Even for the Ly$\alpha$ line considered here, for which the 
		separation between the two interfering $J$-levels is 
		particularly small if compared to that of other FS 
		multiplets, fields of about 2~kG are needed for producing 
		this kind of effects.}
\end{enumerate}

In summary, when the energy separation between pairs of magnetic sublevels 
pertaining to different $J$-levels is much larger than their width (which is 
always the case in the absence of magnetic fields or in the presence 
of weak magnetic fields in the Zeeman effect regime), the contribution of the
interferences between such magnetic sublevels modifies the fractional linear 
polarization pattern in the wings of the lines (outside a $\Delta \lambda 
\simeq 5 \Delta \lambda_D$ wavelength interval around the line center).
However, at these wavelengths the line contribution to the total emissivity 
and absorptivity is no longer dominating with respect to that of the continuum.
The presence of the continuum strongly masks the effects of interferences 
between different $J$-levels, which indeed are observable only in rather
strong spectral lines (e.g., Ca~{\sc ii} H and K, Na~{\sc i} D$_1$ and D$_2$).
Because of the crucial role of the continuum, it is not possible to establish a simple quantitative criterion for deciding whether or not interferences between 
different $J$-levels are expected to produce observable effects.
On the other hand, in the presence of a magnetic field sufficiently intense 
for the Paschen-Back effect regime to be reached (or, in other words, when the 
energy separation between pairs of magnetic sublevels pertaining to different 
$J$-levels is no longer negligible with respect to their width), interferences 
between different $J$-levels may modify the amplitude of the fractional 
polarization pattern in the core of the lines through the mechanisms described
in detail in Sections 10.18 and 10.20 of LL04.

We conclude pointing out that all the results presented in this paper on 
the effects of interferences between different $J$-levels can be generalized to 
the case of interferences between different HFS $F$-levels.
The only remarkable difference for practical applications is the fact that the 
Paschen-Back effect regime for HFS is generally reached for magnetic fields 
much weaker than those needed for reaching the Paschen-Back effect regime for 
FS.

\acknowledgments
{\bf Acknowledgments} Financial support by the Spanish Ministry of Science and 
Innovation through project AYA2010-18029 (Solar Magnetism and Astrophysical 
Spectropolarimetry) is gratefully acknowledged.

\appendix

\section{Analytical results}
In this Appendix, we first recall some equations and analytical results derived 
in LL04 that have been mentioned in the text. 
These equations will be the starting point for the following derivation of a 
series of analytical expressions that will allow us to get more insights on 
some results presented in this paper on the $^2S-{^2P}$ multiplets.
 
Let us consider a two-term atom, the upper and lower terms being characterized 
by the spin $S$ and by the total orbital angular momentum $L$. 
Each term is composed of $(L+S-|L-S|+1)$ FS $J$-levels, each 
having $(2J+1)$ magnetic sublevels $M$, for a total of $(2S+1)\times(2L+1)$ 
sublevels.
As shown in Section~10.16 of LL04, under the simplifying hypotheses of 
unpolarized lower term, no stimulation effects, no collisions, no magnetic 
field, and in the flat-spectrum approximation, it is possible to find an 
analytical solution of the SEEs for the multipole moments of the density matrix 
of the upper term (see Equation~(10.126) of LL04).
If this solution is substituted into the expression of the emission 
coefficient of a two-term atom (see Equation~(7.47e) of LL04), one obtains 
(see Equation~(10.129) of LL04)
\begin{eqnarray}
	\label{Eq:eps2T}
	\varepsilon_i(\nu,\mathbf{\Omega}) & \!\!\! = \!\!\! & 
	\frac{h \nu_0}{4 \pi} \; \mathbb{N}_{\ell} \; 
	B(L_{\ell} \rightarrow L_u) \; \frac{2L_u +1}{2S +1}
	\sum_{KQ} \, \sum_{J_u J_u^{\prime} J_{\ell}} 
	(-1)^{S - L_{\ell} + J_u + J_u^{\prime} +J_{\ell} + K + Q} \nonumber \\
	& & \times \, 3 \, (2J_u +1) \, (2J_u^{\prime}+1) \, (2J_{\ell}+1) 
	\Bigg\{ \!\!
	\begin{array}{ccc} 
		L_u      & L_{\ell} & 1 \\
		J_{\ell} & J_u      & S
	\end{array}
	\!\! \Bigg\} \,
	\Bigg\{ \!\!
	\begin{array}{ccc} 
		L_u      & L_{\ell}     & 1 \\
		J_{\ell} & J_u^{\prime} & S
	\end{array}
	\!\! \Bigg\} \nonumber \\
	& & \times \, 
	\Bigg\{ \!\!
	\begin{array}{ccc} 
		1   & 1            & K        \\
		J_u & J_u^{\prime} & J_{\ell}
	\end{array}
	\!\! \Bigg\} \,
	\Bigg\{ \!\!
	\begin{array}{ccc} 
		1   & 1   & K        \\
		L_u & L_u & L_{\ell}
	\end{array}
	\!\! \Bigg\} \,
	\Bigg\{ \!\!
	\begin{array}{ccc} 
		L_u & L_u          & K   \\
		J_u & J_u^{\prime} & S 
	\end{array}
	\!\! \Bigg\} \nonumber \\
	& & \times \, \mathcal{T}^K_Q(i,\mathbf{\Omega}) \; J^K_{-Q}(\nu_0) \;
	\frac{1}{2} \, \frac{\Phi(\nu_{J_u,J_{\ell}} - \nu) + 
	\Phi(\nu_{J_u^{\prime},J_{\ell}} - \nu)^{\ast}}
	{1 + 2\pi{\rm i} \nu_{J_u^{\prime},J_u}/A(L_u \rightarrow L_{\ell})} 
	\;\; ,  \;\; i=(I,Q,U,V) \;\; ,
\end{eqnarray}
where ${\mathbb N}_{\ell}$ is the number density of atoms in the lower term,
$A(L_u \rightarrow L_{\ell})$ and $B(L_{\ell} \rightarrow L_u)$ are the 
Einstein coefficients for spontaneous emission and for absorption, 
respectively, between the two terms, $\mathcal{T}^K_Q(i,\mathbf{\Omega})$ is a 
geometrical tensor that depends on the direction of the emitted radiation 
$\mathbf{\Omega}$, and on the reference direction for positive $Q$ (see 
Section~5.11 of LL04 and equations therein for its explicit expression),
$\nu_{J_u,J_u^{\prime}}$ is the Bohr frequency between levels $J_u$ and 
$J_u^{\prime}$, and $J^K_Q(\nu_0)$ is the radiation field tensor introduced in 
Section~\ref{Sect:model}, describing the incident (pumping) radiation field. 
We recall that because of the flat-spectrum approximation it is sufficient to 
calculate it at a single frequency $\nu_0$ within the frequency interval 
covered by the multiplet.
The profile $\Phi(\nu_{u \ell} - \nu)$ is given by
\begin{equation}
	\Phi(\nu_{u \ell}-\nu)=\phi(\nu_{u \ell}-\nu)+{\rm i} \, 
	\psi(\nu_{u \ell}-\nu) \;\; ,
\end{equation}
where
\begin{eqnarray}
	\label{Eq:Voigt}
	\phi(\nu_{u \ell}-\nu) & \!\!\! = \!\!\! & \frac{1}{\sqrt{\pi} 
	\Delta \nu_{\rm D}} H(u,a) \;\; , \\
	\label{Eq:Dispersion}
	\psi(\nu_{u \ell}-\nu) & \!\!\! = \!\!\! & \frac{1}{\sqrt{\pi} 
	\Delta \nu_{\rm D}} L(u,a) \;\; .
\label{Eq:profiles}
\end{eqnarray}
The Voigt function $H(u,a)$ and the associated dispersion profile 
$L(u,a)$ are functions of the reduced frequency 
$u \!\! = \!\! (\nu_{u \ell}-\nu)/\Delta \nu_{\rm D}$, with 
$\Delta \nu_{\rm D}$ the Doppler width in frequency units and $\nu_{u \ell}$ 
the Bohr frequency between levels $u$ and $\ell$, and of the damping constant 
$a=\Gamma_{u \ell}/\Delta \nu_{\rm D}$. 
The broadening constant $\Gamma_{ul}$ is given by
\begin{equation}
	\Gamma_{u \ell}=\frac{\gamma_u + \gamma_{\ell}}{4 \pi} \;\; ,
\label{Eq:gamma}
\end{equation}
where $\gamma_u$ and $\gamma_{\ell}$ are the inverse lifetimes of the levels 
$u$ and $\ell$, respectively.

The contribution of interferences between different $J$-levels to the emission 
coefficient is described by the terms with $J_u \ne J_u^{\prime}$ in 
Equation~(\ref{Eq:eps2T}).
If these terms are neglected, recalling the relation
\begin{equation}
	\label{Eq:BEinst}
	B(J_{\ell} \rightarrow J_u) = (2L_{\ell}+1)(2J_u+1)
	\Bigg\{ \!\! 
	\begin{array}{ccc}
		L_u      & L_{\ell} & 1 \\
		J_{\ell} & J_u      & S 
	\end{array}
	\!\! \Bigg\}^2 \;
	B(L_{\ell} \rightarrow L_u) \;\; ,
\end{equation}
and introducing the quantity (see Equation~(10.147) of LL04)
\begin{eqnarray}
	W_K(L_{\ell} L_u S, J_{\ell} J_u) & \!\!\! = \!\!\! &
	(-1)^{S-L_{\ell}-J_{\ell}+K} \; 3 (2L_u+1) (2J_u+1) \nonumber \\
	& & \times 
	\Bigg\{ \!\! 
	\begin{array}{ccc}
		1   & 1   & K \\
		J_u & J_u & J_{\ell}  
	\end{array}
	\!\! \Bigg\} \;
	\Bigg\{ \!\! 
	\begin{array}{ccc}
		1   & 1   & K \\
		L_u & L_u & L_{\ell}  
	\end{array}
	\!\! \Bigg\} \;
	\Bigg\{ \!\! 
	\begin{array}{ccc}
		L_u & L_u & K \\
		J_u & J_u & S 
	\end{array}
	\!\! \Bigg\} \;\; ,
	\label{Eq:WK-gen}
\end{eqnarray}
Equation~(\ref{Eq:eps2T}) reduces to
\begin{eqnarray}
	\label{Eq:epsML}
	\Big[ \varepsilon_i(\nu,\mathbf{\Omega}) \Big]_{\rm no\,int.} 
	& \!\!\! = \!\!\! & 
	\frac{h \nu_0}{4 \pi} \; \mathbb{N}_{\ell} \; \frac{1}{2S+1} \;
	\frac{1}{2L_{\ell}+1} \; \sum_{J_{\ell} J_u} \, (2J_{\ell}+1) \; 
	B(J_{\ell} \rightarrow J_u) \; \phi(\nu_{J_u,J_{\ell}}-\nu) \\ \nonumber
	& & \sum_{KQ} \, W_K(L_{\ell} L_u S, J_{\ell} J_u) \; (-1)^Q \;
	\mathcal{T}^K_Q(i,\mathbf{\Omega}) \; J^K_{-Q}(\nu_0) \;\; .
\end{eqnarray}
This is the expression of the emission coefficient for the corresponding 
multi-level atom, under the same hypotheses required for 
Equation~(\ref{Eq:eps2T}) to hold.
Indeed, it could be obtained starting from the SEEs and from the expression of 
the emission coefficient of a multi-level atom, under the hypotheses of no 
magnetic field, no collisions, no stimulation effects, and assuming that the 
incident radiation field is spectrally flat across all the transitions 
considered and that all the magnetic sublevels of the lower $J$-levels are 
equally populated (hypothesis equivalent to that of the unpolarized lower 
term).

In the particular case of $S=0$, the multiplet reduces to a single line with 
$J_u=L_u$ and $J_{\ell}=L_{\ell}$, the symbol 
$W_K(L_{\ell} L_u S, J_{\ell} J_u)$ reduces to (see Section~10.17 of LL04)
\begin{equation}
	W_K(J_{\ell},J_u)= 3 (2J_u +1)
	\Bigg\{ \!\!
	\begin{array}{ccc}
		1 & 1 & K \\
		J_u & J_u & J_{\ell} 
	\end{array}
	\!\! \Bigg\}^2 \;\; ,
\end{equation}
and the expression of the emission coefficient of a two-level atom is recovered 
(see Equation~(10.16) of LL04):
\begin{equation}
	\label{Eq:eps2L}
	\Big[ \varepsilon_i(\nu,\mathbf{\Omega}) \Big]_{\rm two-lev} = 
	\frac{h \nu_0}{4 \pi} \; 
	\mathbb{N}_{\ell} \; B(J_{\ell} \rightarrow J_u) \; 
	\phi(\nu_{J_u,J_{\ell}}-\nu) \; \sum_{KQ} \, W_K(J_{\ell}, J_u) 
	\;(-1)^Q \; \mathcal{T}^K_Q(i,\mathbf{\Omega}) \; J^K_{-Q}(\nu_0)\;\; .
\end{equation}
The quantity $W_K(L_{\ell} L_u S, J_{\ell} J_u)$ is thus a sort of 
generalization of $W_K(J_{\ell},J_u)$ that takes into account (though under the 
flat-spectrum approximation) the effects due to the presence of the various
components of a multiplet on the polarization properties of a given transition.
Besides the trivial case of $S=0$, it can be shown that the symbol
$W_K(L_{\ell} L_u S, J_{\ell} J_u)$ coincides with the symbol 
$W_K(J_{\ell},J_u)$ in all the $S - P$ multiplets (all the multiplets with 
$L_{\ell}=0$, $L_u=1$, and any value of the spin).

We consider an unmagnetized plane-parallel atmosphere, we take a Cartesian 
reference system with the $z$-axis (quantization axis) directed along the 
local vertical and focus our attention on the radiation scattered at 
90$^{\circ}$ by an optically thin slab of solar plasma.
Choosing the reference direction for positive $Q$ parallel to the atmosphere,
and recalling that, because of the symmetry of the problem, the only non-zero 
components of the radiation field tensor are $J^0_0$ and $J^2_0$, the 
geometrical tensors $\mathcal{T}^K_Q(i,\mathbf{\Omega})$ that enter the 
previous expressions of the emission coefficients assume the values
\begin{eqnarray}
	\mathcal{T}^0_0(I,\mathbf{\Omega}) = 1 \;\; , & \;\; \;\;\; &
	\mathcal{T}^2_0(I,\mathbf{\Omega}) = -\frac{1}{2 \sqrt{2}} \;\; , \\
	\mathcal{T}^0_0(Q,\mathbf{\Omega}) = 0 \;\; , & \;\; \;\;\; &
	\mathcal{T}^2_0(Q,\mathbf{\Omega}) = \frac{3}{2 \sqrt{2}} \;\; .
\end{eqnarray}

Substituting these values into Equation~(\ref{Eq:eps2L}), we can easily obtain 
the analytical expression of the fractional polarization pattern 
$p_Q(\nu) = \varepsilon_Q(\nu)/\varepsilon_I(\nu)$ for the radiation scattered 
at 90$^{\circ}$ by a two-level atom:
\begin{equation}
	\Big[ p_Q(\nu) \Big]_{\rm two-lev} = \frac{3W_2(J_{\ell},J_u)}
	{\displaystyle \frac{4}{w} - W_2(J_{\ell},J_u)} \;\; ,
\label{Eq:pq-2L}
\end{equation}
where $w$ is the anisotropy factor defined in Equation~(\ref{Eq:w-barn}).

As discussed in LL04, in a two-term atom, when the separation among the various 
lines of the multiplet is much larger than their natural width, the emission 
coefficients in the neighborhood of a single line with $J_u = \bar{J}_u$ and 
$J_{\ell} = \bar{J}_{\ell}$ can be evaluated by restricting the summation over 
$J_u$, $J_u^{\prime}$, and $J_{\ell}$ in Equation~(\ref{Eq:eps2T}) to the 
values $J_u = J_u^{\prime} = \bar{J}_u$ and $J_{\ell} = \bar{J}_{\ell}$ (i.e.,
neglecting, in particular, the terms corresponding to the interferences between 
different $J$-levels).
From Equation~(\ref{Eq:epsML}), we can thus obtain the analytical expression of 
the fractional polarization of the radiation scattered by a two-term atom in 
the core of the various lines of the corresponding multiplet.
For 90$^{\circ}$ scattering we obtain
\begin{equation}
	\Big[ p_Q({\rm core}) \Big]_{\rm two-term} = \frac{3W_2(L_{\ell} L_u S, 
	J_{\ell} J_u)}
	{\displaystyle \frac{4}{w} - W_2(L_{\ell} L_u S, J_{\ell} J_u)} \;\; .
\label{Eq:pq-2Tcore}
\end{equation}

We now focus our attention on the $^2S - ^2P$ multiplet. Taking into account 
that for this multiplet $W_K(L_{\ell}L_u S,J_{\ell}J_u)\!=\!W_K(J_{\ell},J_u)$ 
and that $W_2(1/2,1/2)\!=\!0$ (transition 1), while $W_2(1/2,3/2)\!=\!1/2$ 
(transition 2), starting from Equations~(\ref{Eq:eps2T}) and (\ref{Eq:epsML}), 
it is possible to find rather compact analytical expressions for the fractional 
polarization pattern $p_Q(\nu)$ of the radiation scattered at 90$^{\circ}$, 
both neglecting or taking into account interferences.
Using the more compact notation $\phi_j \equiv \phi(\nu_{0j}-\nu)$ and 
$\psi_j \equiv \psi(\nu_{0j}-\nu)$ for the profiles defined in 
Equation~(\ref{Eq:profiles}), with $\nu_{0j}$ the frequency of transition $j$, 
after some algebra (involving the calculation of several 6-$j$ symbols, and the 
use of Equation~(\ref{Eq:BEinst})), one obtains the following expressions:\\
\begin{equation}
	\Big[ p_Q(\nu) \Big]_{\rm no\,int.} = \frac{3/2 \, \phi_2}
	{\displaystyle \frac{4}{w} \, \Bigg( \frac{1}{2} \phi_1 + \phi_2 \Bigg)
	- \frac{1}{2}\,\phi_2} \;\; ,
\label{Eq:eps3Lp}
\end{equation}
\begin{equation}
	\Big[ p_Q(\nu) \Big]_{\rm int.} = \frac{3/2 \, (\phi_2 + \chi)}
	{\displaystyle \frac{4}{w} \, \Bigg( \frac{1}{2} \phi_1 + \phi_2 \Bigg)
	-\frac{1}{2} \Bigg( \phi_2 + \chi \Bigg)} \;\; ,
\label{Eq:eps2Tp}
\end{equation}
where
\begin{equation}
	\chi=\frac{1}{1+\alpha^2} \, (\phi_1+\phi_2) + 
	\frac{\alpha}{1+\alpha^2} \, (\psi_1-\psi_2) \;\; ,
\label{Eq:chi}
\end{equation}
with $\alpha = (2 \pi \Delta \nu)/A(L_u \rightarrow L_{\ell})$. 
The quantity $\Delta \nu = \nu_{02}-\nu_{01}$ is the frequency separation 
between the two components of the multiplet.
As it can be observed comparing Equations~(\ref{Eq:eps3Lp}) and 
(\ref{Eq:eps2Tp}), the contribution of interferences between different 
$J$-levels is fully described by the quantity $\chi$.

\begin{figure}[!t]
\includegraphics[width=\textwidth]{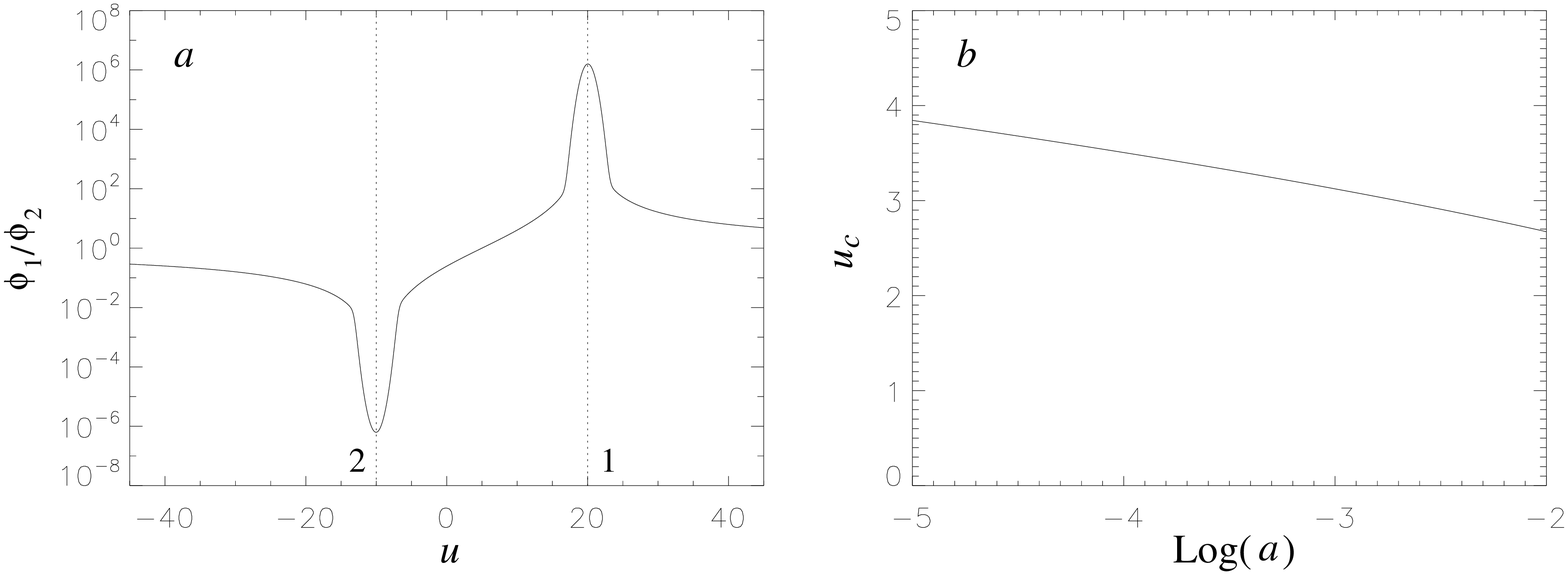}
\caption{{\footnotesize Panel $a$: ratio between the Voigt profile of 
transition 1 ($\phi_1$), and the Voigt profile of transition 2 ($\phi_2$), 
plotted as a function of the reduced wavelength 
$u=(\lambda-\lambda_0)/\Delta \lambda_D$, with $\Delta \lambda_D$ the Doppler 
width of the two lines, and $\lambda_0$ the wavelength corresponding to the 
energy difference between the centers of gravity of the two terms.
The separation between the two components is $\Delta \lambda = 30 \,\Delta 
\lambda_D$. The damping parameter is $a=10^{-3}$.
Panel $b$: plot of $u_c$ (see the text) as a function of Log($a$).}}
\label{Fig:Voigt-vc}
\end{figure}

In Section~\ref{Sect:DW-effect}, we observed that in the neighborhood of the 
single transitions, the $p_Q(\nu)$ profiles calculated taking into account 
and neglecting interferences coincide and are constant over a 
wavelength interval of about five Doppler widths.
The analytical expressions of $p_Q$ given by Equations~(\ref{Eq:eps3Lp}) and 
(\ref{Eq:eps2Tp}) allow to analyze in detail the origin of this behavior.
As far as $[ p_Q(\nu) ]_{\rm no\,int.}$ is concerned, 
Equation~(\ref{Eq:eps3Lp}) can be rewritten as
\begin{equation}
	\Big[ p_Q(\nu) \Big]_{\rm no\,int.} = \frac{3/2}
	{\displaystyle \frac{4}{w} \, \Bigg(1+\frac{1}{2} \frac{\phi_1}{\phi_2} 
	\Bigg) - \frac{1}{2}} \;\; .
\end{equation}
As shown in panel~$a$ of Figure~\ref{Fig:Voigt-vc}, the ratio $\phi_1/\phi_2$ 
is much smaller than unity for frequencies close to transition 2, while it is 
extremely large for frequencies close to transition 1.
In the core of the two transitions, we thus obtain
\begin{equation}
	\Big[ p_Q(\nu \approx \nu_{02}) \Big]_{\rm no\,int.} = \frac{3/2}
	{\displaystyle \frac{4}{w} - \frac{1}{2}} \;\; ,
	\;\;\;\;\;\;\;
	\Big[ p_Q(\nu \approx \nu_{01}) \Big]_{\rm no\,int.} = 0 \;\; .
\end{equation}
These expressions coincide with those valid for a two-level atom
(see Equation~(\ref{Eq:pq-2L})).

If the two lines are very separated from each other, the boundaries of the 
intervals over which the ratio $\phi_1/\phi_2$ is very small or very large
(and thus the fractional polarization $[ p_Q(\nu) ]_{\rm no \, int.}$ is 
practically constant) are determined by the points where the shape of the 
Voigt profile changes from Gaussian to Lorentzian (see panel $a$ of 
Figure~\ref{Fig:Voigt-vc}).

Exploiting a series of properties of the functions $H(u,a)$ and 
$L(u,a)$ discussed in Section~5.4 of LL04, it can be shown that an asymptotic 
expansion of the Voigt function in power series of $a$ is given by
\begin{equation}
	H(u,a) \simeq {\rm e}^{-u^2} + \frac{1}{\sqrt{\pi}} 
	\frac{a}{u^2} \;\; .
\end{equation}
The value $u_c$ for which the behavior of the Voigt function changes from 
Gaussian to Lorentzian can thus be evaluated through the equality
\begin{equation}
	{\rm e}^{-u_c^2} = \frac{1}{\sqrt \pi} \frac{a}{u_c^2} \;\; .
\end{equation}
It can be verified that the solution of this transcendent equation is 
given by the following recursive expression (E. Landi Degl'Innocenti, 
private communication):
\begin{equation}
	u_c=\sqrt{\ln(C \, \ln(C \, \ln(C \, \ln(C\, \ln(C\, \cdots)))))} 
	\;\; ,
\end{equation}
with $C=\sqrt{\pi}/a$.
The value of $u_c$ as a function of Log($a$) is shown in panel~$b$ of 
Figure~\ref{Fig:Voigt-vc}. 
We see that $u_c$ varies almost linearly with Log($a$), going 
from a value of 3.844 for $a=10^{-5}$ to a value of 2.673 for $a=10^{-2}$, 
in agreement with the extension (of about five Doppler widths) of the plateaux 
observed in the plots of Section~\ref{Sect:DW-effect}.
\begin{figure}[!t]
\includegraphics[width=\textwidth]{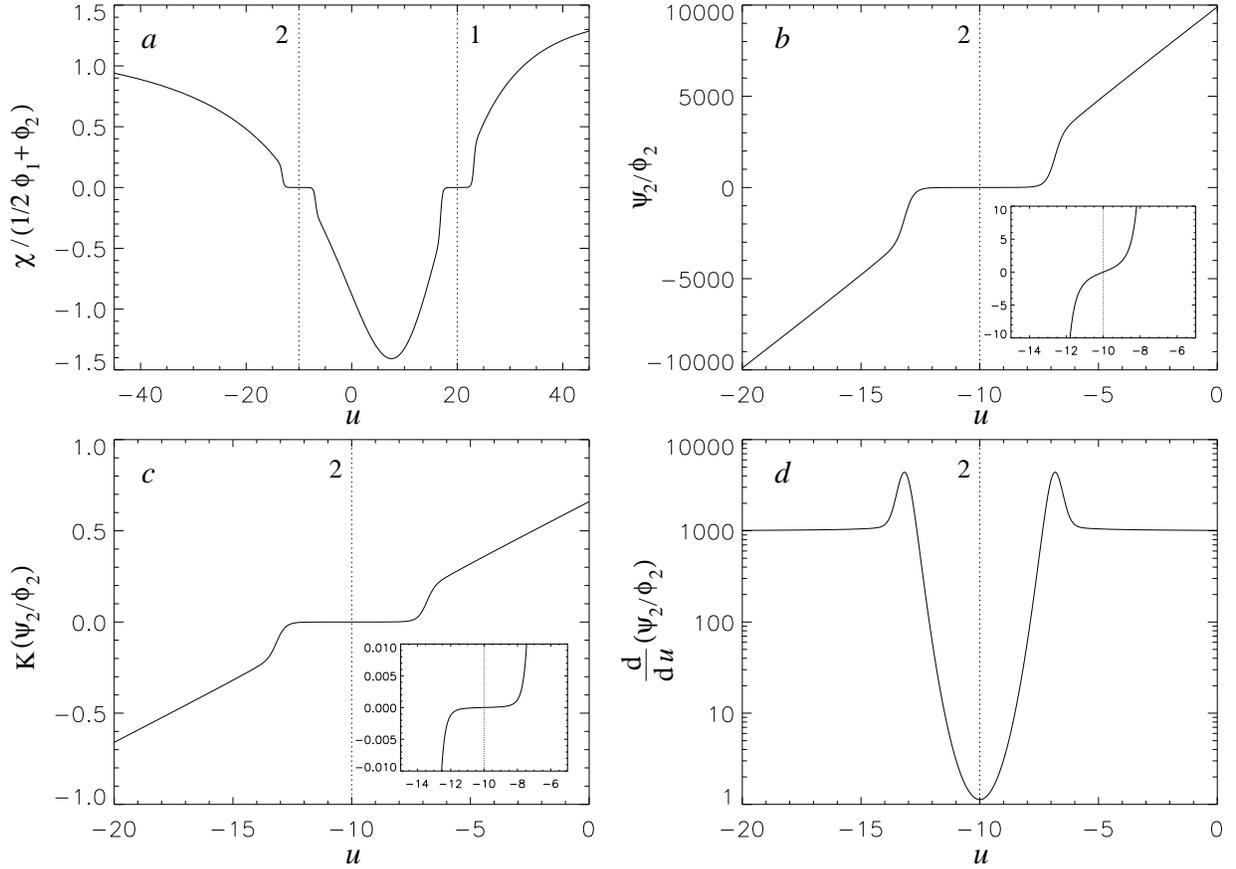}
\caption{{\footnotesize Panel $a$: plot of the interference term 
$\chi/(1/2 \phi_1+ \phi_2)$ as a function of the reduced wavelength 
\mbox{$u=(\lambda-\lambda_0) / \Delta\lambda_D$}, with $\Delta \lambda_D$ the 
Doppler width of the two lines, and $\lambda_0$ the wavelength corresponding 
to the energy difference between the centers of gravity of the two terms.
As in panel $a$ of Figure~\ref{Fig:Voigt-vc}, the separation between the two 
components is $\Delta \lambda = 30 \, \Delta \lambda_D$, while the damping 
parameter is $a=10^{-3}$.
Panel $b$: ratio between the dispersion profile of line 2 ($\psi_2$) and the 
Voigt profile of line 2 ($\phi_2$). The inner panel shows in more details the 
behavior of this ratio around the wavelength position of transition 2. 
Panel $c$: Same as panel $b$, with the ratio $\psi_2/\phi_2$ multiplied by 
$K=\alpha/(1+\alpha^2)=6.67 \times 10^{-5}$. 
Panel $d$: Plot of the first derivative of the ratio $\psi_2/\phi_2$.}}
\label{Fig:chi-Voigt}
\end{figure}

The situation is similar as far as the $[ p_Q(\nu) ]_{\rm int.}$ 
profile is concerned.
Noticing that for small values of $w$ the second term in the denominator of
Equation~(\ref{Eq:eps2Tp}) is much smaller than the first one, this equation 
can be simplified as
\begin{equation}
	\Big[ p_Q(\nu) \Big]_{\rm int.} = \frac{3w}{8} \left[
	\frac{\phi_2}{\frac{1}{2} \phi_1 + \phi_2} +
	\frac{\chi}{\frac{1}{2} \phi_1 + \phi_2} \right] \;\; .
\end{equation}
The contribution of interferences is described by the term 
$\chi / (\frac{1}{2}\phi_1 + \phi_2)$ (see panel~$a$ of 
Figure~\ref{Fig:chi-Voigt}), the remaining part of the expression being the 
same as in the case without interferences.
In order to analyze the behavior of this term around transition 2,
recalling Equation~(\ref{Eq:chi}), we rewrite it as
\begin{equation}
	\frac{\chi}{\frac{1}{2} \phi_1 +\phi_2} =
	\frac{1}{1 + \frac{1}{2} \frac{\phi_1}{\phi_2}}
	\Bigg[ \frac{1}{1+\alpha^2} \Bigg( 1+\frac{\phi_1}{\phi_2} \Bigg) +
	\frac{\alpha}{1+\alpha^2} \Bigg( \frac{\psi_1}{\phi_2} - 
	\frac{\psi_2}{\phi_2} \Bigg) \Bigg] \;\; .
	\label{Eq:interf}
\end{equation}
Noticing that when the two transition are sufficiently separated from each 
other, as we are assuming here, $\phi_1$ and $\psi_1$ are practically constant 
close to transition 2, it is clear that both the ratios $\phi_1/\phi_2$ (as 
observed in panel $a$ of Figure~\ref{Fig:Voigt-vc}) and $\psi_1/\phi_2$ are 
practically zero in the frequency interval over which $\phi_2$ has a Gaussian 
behavior.
More complex is the behavior of $\psi_2/\phi_2$.
This ratio is exactly zero for $u=0$ but, as shown in panel~$b$ of 
Figure~\ref{Fig:chi-Voigt}, it immediately increases (in absolute value) moving 
away from the line center, assuming values of the order of unity already for 
$u \approx 1$.
The first derivative of this ratio shows an abrupt increase at the frequencies 
where the behavior of the Voigt function $\phi_2$ changes from Gaussian to 
Lorentzian, becoming practically constant for larger values of $u$ 
(see panel $d$ of Figure~\ref{Fig:chi-Voigt}).
However, as shown in panel~$c$ of Figure~\ref{Fig:chi-Voigt}, the contribution 
of this term is in any case negligible within the interval where the Voigt 
function $\phi_2$ has a Gaussian behavior, when the multiplicative factor 
$K=\alpha/(1+\alpha^2)$ (which is extremely small when the separation 
between the two components is much larger than the Einstein coefficient) 
is taken into account.
Noticing that the multiplicative factor $1/(1+\alpha^2)$ makes the contribution 
of the first term in the square bracket of Equation~(\ref{Eq:interf}) 
negligible within the interval over which $\phi_2$ has a Gaussian behavior, it 
follows that interferences bring a negligible contribution on this spectral 
interval around transition 2.

With analogous considerations, it can be shown that interferences bring a 
negligible contribution around transition 1, within the interval over 
which the Voigt profile $\phi_1$ has a Gaussian behavior.

\end{document}